\documentclass[11pt]{iopart}
\usepackage{color}
\usepackage{graphicx}
\usepackage{psfrag}
\usepackage{times}
\usepackage{iopams}
\usepackage{textcomp}

\usepackage{dsfont}
\usepackage{bm}
\usepackage{mathrsfs}

\usepackage{hyperref}

\usepackage[T1]{fontenc}
\usepackage{textcomp} 
\usepackage[latin1]{inputenc}
\usepackage{fixmath} 
\usepackage{exscale}

\usepackage[dcucite]{harvard}
\citationmode{abbr}

\newcommand{\Mi}{\rmi} 
\newcommand{\Me}{\rme} 
\newcommand{\Mbf}[1]{\bi{#1}} 
\newcommand{\Mdiff}{\mathrm{d}} 
\newcommand{\Mlabel}[1]{\mathrm{#1}} 

\newcommand{\eqref}[1]{(\ref{#1})}
\newcommand{\op}[1]{{\rm #1}}
\renewcommand{\tr}[1]{{{\rm Tr}\left\{#1\right\}}}
\newcommand{\tp}{{\rm T}} 
\newcommand{\id}{\mathds{1}}

\newcommand{\braket}[2]{\left\langle#1|#2\right\rangle}
\newcommand{\abs}[1]{\left|#1\right|}

\newcommand{\ave}[1]{\left\langle#1\right\rangle}
\newcommand{\dd}{\partial}
\newcommand{\diff}[2]{\frac{\Mdiff#1}{\Mdiff#2}}
\newcommand{\pdiff}[2]{\frac{\dd#1}{\dd#2}}
\newcommand{\diffz}[2]{\frac{\Mdiff^2#1}{\Mdiff#2^2}}

\renewcommand{\brho}{\boldsymbol{\rho}}
\renewcommand{\bzeta}{\boldsymbol{\xi}} 
\newcommand{\rr}{\Mbf{x}} 
\newcommand{\mboxbf}[1]{\Mbf{#1}} 
\newcommand{\bpp}{\hat\Mbf{p}}

\newcommand{\bgg}{\Mbf{F}}
\newcommand{\calR}{\boldsymbol{\mathcal{R}}}
\newcommand{\calP}{\boldsymbol{\mathcal{P}}}

\newcommand{\calS}{\mathcal{S}}
\newcommand{\calL}{\mathcal{L}}
\newcommand{\GammaF}{\mathrm{\Gamma}}
\providecommand{\tfrac}[2]{{\textstyle\frac{#1}{#2}}}
\newcommand{\das}{\equiv}

\newcommand{\BS}{\begin{eqnarray}}
\newcommand{\ES}{\end{eqnarray}}
\newcommand{\BSW}{\begin{eqnarray*}}
\newcommand{\ESW}{\end{eqnarray*}}
\newcommand{\R}{\mathds{R}}
\newcommand{\Reals}{\R}

\newcommand{\Zero}{0}
\newcommand{\Unit}{\mathds{1}}
\providecommand{\bx}{\Mbf{x}}

\begin{document}
\frenchspacing

\title{Efficient description of Bose-Einstein condensates in time-dependent
rotating traps}

\author{M.~Meister$^1$, S.~Arnold$^2$, D.~Moll$^1$, M.~Eckart$^1$, E.~Kajari$^1$, 
M.~A.~Efremov$^{1}$, R.~Walser$^3$, W.~P.~Schleich$^{1,4}$}

\address{$^1$ Institut f\"ur Quantenphysik and Center for Integrated Quantum Science and Technology (IQ$^\mathrm{ST}$),
Universit\"at Ulm, D-89081 Ulm, Germany}
\address{$^2$ Institut f\"ur Theoretische Informatik, Universit\"at Ulm,
D-89081 Ulm, Germany}
\address{$^3$ Institut f\"ur Angewandte Physik, Technische Universit\"at Darmstadt,
D-64289 Darmstadt, Germany}
\address{$^4$ Hagler Institute for Advanced Study at Texas A\&M University, Institute for Quantum Science and Engineering (IQSE), and Department of Physics and Astronomy, Texas A\&M University, College Station, TX 77843-4242, USA}

\ead{matthias.meister@uni-ulm.de}

\begin{abstract}
Quantum sensors based on matter-wave interferometry are promising candidates for high-precision gravimetry and inertial sensing in space. The favorable source for the coherent matter waves in these devices are Bose-Einstein condensates. A reliable prediction of their dynamics, which is governed by the Gross-Pitaevskii equation, requires suitable analytical and numerical methods which take into account the center-of-mass motion of the condensate, its rotation and its spatial expansion by many orders of magnitude. 

In this chapter, we present an efficient way to study their dynamics in time-dependent rotating traps that meet this objective. Both, an approximate analytical solution for condensates in the Thomas-Fermi regime and dedicated numerical simulations on a variable adapted grid are discussed. We contrast and relate our approach to previous alternative methods and provide further results, such as analytical expressions for the one- and two-dimensional spatial density distributions and the momentum distribution in the long-time limit that are of immediate interest to experimentalists working in this field of research.
\\ \\ 
\vspace{1cm}
\textbf{Date:} July 7th, 2017
\\ \\ \vfill\hfill
Preprint submitted to Elsevier 
\end{abstract}

\maketitle

\section{Introduction}
Since the first creation of a Bose-Einstein condensate (BEC) in the mid 1990s \cite{Anderson95,Davis95}, the field of ultra-cold quantum gases \cite{DGPS99,Giorgini2008} has enormously developed and BECs are nowadays commonly used in a broad variety of applications. They include the generation of vortices \cite{Fetter2009}, the exploration of different quantum phase transitions \cite{Bloch2008} as well as inertial sensors that are realized with the help of matter-wave interferometry \cite{Berman1997,Cronin2009,Tino2014}. In many of these phenomena the Gross-Pitaevskii (GP) equation \cite{Gross61,Gross63,Pitaevskii61} provides a reliable theoretical description of the BEC dynamics. Analytic solutions of the GP equation can be derived within the Thomas-Fermi (TF) approximation when combined with the so-called scaling approach \cite{kagan96,castin96,kagan97}, where most of the BEC dynamics is described by an appropriate time-dependent coordinate transformation. In this chapter we generalize this scaling 
approach and apply it directly to the GP equation to obtain an efficient analytic description of the dynamics of a BEC in time-dependent rotating traps. 

Our approach was motivated and used by seminal experiments realized within the QUANTUS collaboration \cite{VanZoest2010,Muentinga2013}, which has successfully performed matter-wave interferometry with BECs in microgravity at the drop tower in Bremen (ZARM). These pioneering experiments study and manipulate BECs that expand freely for several seconds and hence evolve into matter waves of macroscopic dimensions. In order to reliably predict and describe the outcome of such experiments, it is essential to have dedicated analytical and numerical tools at hand which take into account translational and rotational motions of the BEC as well as the fact that the spatial size of the condensate changes by many orders of magnitude during its free expansion. 
To account for these effects, the scaling approach has been generalized within the hydrodynamic framework \cite{storey00,Edwards2002} by employing an affine transformation, where a translation first eliminates the center-of-mass motion and a subsequent linear map absorbs most of the remaining dynamics of the BEC. Here we apply this affine transformation directly to the GP equation resulting in an efficient description of the time evolution of a BEC in a time-dependent quadratic potential including slow rotations. Our approach has a straightforward application to matter-wave interferometry \cite{Roura2014}, facilitates an efficient numerical computation of the condensate wave function and provides valuable analytical insights into the dynamics of BECs.

Our chapter is organized as follows. In section \ref{sec:efficient_description} we first present the affine transformation of the GP equation, where we displace the wave function to eliminate the center-of-mass motion and then apply a linear transformation of the coordinates to account for the inner dynamics of the BEC. The combination of the affine approach and the time-dependent TF approximation results in an approximate, but astonishingly accurate solution for the dynamics of a BEC subject to a time-dependent rotating trap. Moreover, we derive expressions for the one- and two-dimensional integrated density distributions, which are experimentally accessible through time-of-flight pictures.

To verify the accuracy of our efficient description of the BEC dynamics, we perform in section \ref{Sec:Numerical} full numerical simulations of the GP equation for a purely rotating trap as well as for the free expansion of an initially rotating BEC. We find an excellent agreement with our approximate analytical solution and show that the affine transformation can be used to improve the performance of numerical simulations by solving the transformed GP equation rather than the original one. Indeed, since the affine transformation itself does not contain any approximation, our technique does not alter the accuracy of the numerical simulations but speeds up the computation. 

In section \ref{chap-constants-of-motion} we establish a connection between the dynamics of the time-dependent affine transformation and the corresponding Hamiltonian formalism triggered by \cite{kagan97}. In this context, we discuss two constants of motion of the affine transformation matrix, which we relate to the conservation of the total energy and the angular momentum of a BEC. 

Since almost every experiment dealing with BECs makes use of time-of-flight pictures, it is essential to have a thorough understanding of the dynamics of a BEC during its free expansion. Indeed, due to their mean-field interaction, BECs posses a more complex time-of-flight dynamics compared with the simple ballistic expansion known from non-interacting quantum gases. Therefore, in section \ref{sec:free-expansion-properties} we compare the free time evolution of the GP equation with the Schr\"odinger equation and derive a relation between the long-time behavior of the momentum distribution of a freely expanding BEC and its initial spatial density distribution.

In order to keep our consideration self-contained but focused on the central ideas, we have included further details and calculations in three appendices. In \ref{AppAffineTrafo} we sketch the derivation of the affinely transformed GP equation and show how to calculate the integrated density distributions of a BEC within the TF regime. The important steps to determine the total energy of a BEC and its angular momentum within the time-dependent TF approximation are presented in \ref{sec:app:constants_of_motion}. Finally, in \ref{app:Lambda:DGL:isotropic} we discuss the behavior of the affine transformation matrix for a freely expanding BEC that is released from an isotropic harmonic trap.

\section{Efficient description of the time evolution of a Bose-Einstein condensate}\label{sec:efficient_description}\label{chap:efficient_description}
The present section introduces a natural generalization of
the scaling approach \cite{kagan96,castin96,kagan97} to the case of {\it rotating
harmonic traps}. In contrast to the hydrodynamical approach \cite{storey00}, 
we carry out this generalization directly to the GP equation. 
We apply these results to a BEC within the TF regime
and obtain an approximate, analytical description of its time evolution, 
the so-called time-dependent TF approximation. Moreover, we then derive
simple expressions for the integrated density distributions of a BEC within the
TF regime which establish a direct link to experimental time-of-flight observations.
We conclude this section with a brief discussion of an alternative, but equivalent 
description~\cite{Edwards2002} of the internal dynamics of a BEC. 
Further details on the results presented in this section are collected in~\ref{AppAffineTrafo}.

\subsection{Affine transformation of the Gross-Pitaevskii equation}
\label{aff_tran}

The affine transformation of the GP equation is realized in two steps \cite{PhD_ME}, namely 
\textit{(i)} we eliminate the center-of-mass motion using Kohn's theorem 
\cite{Kohn1961,Dobson1994,Bialynicki-Birula2002}, and
\textit{(ii)} we introduce a linear mapping of the coordinates in order to account for the main 
contributions to the internal dynamics of a BEC.
Moreover, we show how this linear mapping is reduced to the well-known scaling approach \cite{kagan96,castin96,kagan97} 
in the case of a non-rotating harmonic potential.

\subsubsection{Basic setting}

We describe the dynamics of a BEC in an inertial frame of reference by the macroscopic wave function $\psi(t,\rr)$ 
which satisfies the GP equation 
\begin{equation}
  \label{GPequation}
  \Mi\hbar\,\pdiff{\psi}{t}
  = \left[-\frac{\hbar^2}{2m}\bnabla^2_{\!\rr}
  + V(t,\rr)
  + g\,|\psi(t,\rr)|^2\right]\psi(t,\rr),
\end{equation}
where the position vector $\rr$ is considered here as an element of the $d$-dimensional vector
space $\mathds{R}^d$ with $d=1,2,{\rm or}\, 3$. In this way, our results are applicable
to the general three-dimensional case \cite{cornell796,ketterle896,Stamper-Kurn1998,DGPS99,Ketterle1999,Hechenblaikner2002,Edwards2002,Clancy2007} as well as to the cases of BECs confined in one or two dimensions
\cite{Petrov2000,Plaja2002,Bongs2001,Goerlitz2001}.
Throughout this article the macroscopic wave function $\psi(t,\rr)$ is normalized to the number of particles $N$ in the condensate, that is
\begin{equation}
  \label{eq:Normalization}
  \int_{\mathds{R}^d} |\psi(t,\rr)|^2\,\Mdiff^d x = N\,.
\end{equation}
The general potential $V(t,\rr)$ in Eq.~\eqref{GPequation} describes the interaction of an atom of mass $m$ with the external fields that correspond for example to a magneto-optical trap and Earth's gravity.
Moreover, we assume that the atoms interact with each other via a repulsive contact interaction, 
leading to the non-linear term $g\,|\psi(t,\rr)|^2$ in Eq. (\ref{GPequation}) with the positive coupling constant $g$.

In many experimental situations a semiclassical treatment of the underlying quantum mechanical dynamics allows an accurate 
and transparent interpretation of the measurement results. Here, we apply this approach to study the BEC dynamics  
around a classical trajectory $\brho(t)$ and expand the potential $V(t,\rr)$ into a Taylor series 
up to second order around this trajectory
\begin{equation}
 \label{eq:potential}
  V(t,\rr)= V(t,\brho(t))- \bgg(t)[\rr-\brho(t)]+\frac{m}{2}[\rr-\brho(t)]^{\tp}\Omega^{2}(t)\,[\rr-\brho(t)]\,.
\end{equation}
Any anharmonicity of the potential can be neglected, as long as the size of the BEC remains sufficiently small within the vicinity of the trajectory $\brho(t)$. The latter can be associated with the center-of-mass motion of the atomic cloud 
or the minimum of the external potential $V(t,\rr)$, as discussed in more detail in section 
\ref{sec:rho}.

Each term in Eq. \eqref{eq:potential} has a clear meaning. Indeed, 
the zeroth order term $V(t,\brho(t))$ represents the value of the potential along the trajectory $\brho(t)$. 
The second term ${-\bgg(t)[\rr-\brho(t)]}$ corresponds to the force $\bgg(t)$ 
acting on the atoms at the point $\rr=\brho(t)$. 
The third term $\frac{1}{2}m[\rr-\brho(t)]^{\!\tp}\Omega^2(t)[\rr-\brho(t)]$ is the purely quadratic trapping potential 
represented in terms of the symmetric positive definite matrix ${\Omega^{2}(t)}$. Its eigenvalues coincide with
the squared trap frequencies $\omega_i^2(t)$ along the principal axes of the harmonic trap.

\subsubsection{Elimination of the center-of-mass motion \label{sec:com_Elimination}}

We incorporate the center-of-mass motion of the BEC in a straightforward way and 
thereby eliminate all $\brho$-dependent terms in the potential~\eqref{eq:potential}
by making use of the transformation 
\begin{equation}
 \psi(t,\rr)
 = \Me^{\frac{\Mi}{\hbar}\calS_1(t)}\,
   \Me^{\frac{\Mi}{\hbar}\left[\calP(t)\,\rr-\calR(t)\,\bpp\right]} \,
   \psi_\Mlabel D(t,\rr)
\label{DefpsiD}
\end{equation} 
from the original wave function $\psi(t,\rr)$ to the new one $\psi_\Mlabel D(t,\rr)$. Note that 
$\bpp\equiv -\Mi\hbar\bnabla_{\rr}$ represents the momentum operator in the position representation.

The time-dependent vectors $\calR(t)$ and $\calP(t)$ in Eq. \eqref{DefpsiD} describe the time evolution of the
center of mass of the condensate, see section~\ref{sec:Ehrenfest} for more details, 
and obey the classical equations of motion
\begin{eqnarray}
 \label{eq:equationsOfMotion}
  \diff{\calR}{t}&=&\frac{\calP(t)}{m}\,,\nonumber\\
  \diff{\calP}{t}&=&-m\Omega^{2}(t)\left[\calR(t)-\brho(t)\right]+\bgg(t),
\end{eqnarray}
which are the Hamilton equations corresponding to the Lagrangian function
\begin{equation}
 \label{eq:Lagrangian}
 \calL (\calR,\dot{\calR},t) =
 \frac{m}{2}\,\dot{\calR}^2-V(t,\brho)+\bgg(\calR-\brho)
 -\frac{m}{2}\,(\calR-\brho)^{\tp}\Omega^2(\calR-\brho)\,.
\end{equation}
The gobal phase $\calS_1(t)$ in Eq.~\eqref{DefpsiD} depends on the classical action via the generalized definition
\begin{equation}
 \label{eq:PhaseS}
 \calS_k(t)= \int_{0}^t \calL(\calR,\dot{\calR},t')\Mdiff t'
 - \frac{k}{2}\left[\calR(t)\calP(t)-\calR(0)\calP(0)\right],
\end{equation} 
where the additional integer $k\in \mathds{Z}$ has been introduced for later purposes.
As outlined in \ref{App_COM}, we arrive at the GP equation 
\begin{equation}
 \label{GPequationD}
 \Mi\hbar\,\pdiff{\psi_\Mlabel D}{t}
 = \left[-\frac{\hbar^2}{2m}\bnabla^2_{\!\rr}
 + \frac{m}{2}\,\rr^{\!\tp}\Omega^{2\!}(t)\,\rr
 + g\,|\psi_\Mlabel D(t,\rr)|^2\right]\psi_\Mlabel D(t,\rr)
\end{equation}
for the transformed wave function $\psi_\Mlabel D(t,\rr)$
by inserting Eq.~\eqref{DefpsiD} into Eq. \eqref{GPequation} and 
taking advantage of Eqs. \eqref{eq:equationsOfMotion}-\eqref{eq:PhaseS} for $k=1$. 

The decoupling of the center-of-mass motion of a BEC is possible as long as the potential is at
most quadratic \cite{Dobson1994,Bialynicki-Birula2002,Nandi07}. It no longer holds true for
anharmonic potentials for which a nontrivial coupling of
the center-of-mass motion and the inner dynamics of the condensate exists \cite{dum98}.

\subsubsection{Linear transformation as a natural generalization of the
scaling approach\label{SecLinearTrafo}}

After having eliminated the center-of-mass motion, we perform a time-dependent linear
transformation to account for the main internal dynamics of the BEC. For this purpose, we also
introduce a new time coordinate $\tau$ such that the linear mapping between the 
``original coordinates'' $(t,\rr)$ and the new ones $(\tau,\bzeta)$ reads
\begin{eqnarray}
t&=&\tau \,,\nonumber\\
\rr&=&\Lambda(\tau)\,\bzeta\,,
\label{lineartrafo}
\end{eqnarray}
with $\Lambda(\tau)$ being an arbitrary, time-dependent matrix. In what follows, we refer 
to $\Lambda(\tau)$ as ``adaptive matrix'' and to $(\tau,\bzeta)$ as ``adapted coordinates''. 

In analogy to Eq. \eqref{DefpsiD}, the coordinate transformation~\eqref{lineartrafo} 
goes hand in hand with the transformation 
\begin{equation}
  \psi_\Mlabel D(t,\rr)
  = \frac{1}{\sqrt{\det \Lambda(\tau)}}\;
  \Me^{\frac{\Mi}{\hbar}\left[\bzeta^\tp A(\tau)\bzeta-\beta(\tau)\right]}\,
  \psi_\Mlabel\Lambda(\tau,\bzeta)
\label{DefpsiLambda}
\end{equation}
to the so-called affinely transformed wave function $\psi_\Mlabel\Lambda(t,\rr)$. 
The scalar phase $\beta(\tau)$ and the symmetric matrix $A(\tau)$ 
introduced in this transformation depend on the adaptive matrix~$\Lambda(\tau)$ via
\begin{equation}
 \label{Def_beta}
  \beta(\tau)\equiv \int_0^\tau \frac{\mu}{\det \Lambda(\tau')}\,\Mdiff\tau'
\end{equation}
and 
\begin{equation}\label{Def_A}
 A(\tau)\equiv \frac{m}{2}\,\Lambda^\tp\!(\tau)\,\diff{\Lambda}{\tau} \,.
\end{equation}
The constant $\mu$ that appears in the definition~\eqref{Def_beta} 
represents the chemical potential associated with the ground state of the initial BEC at the time $\tau=0$.
In \ref{App_Lambda} we outline the derivation of the affinely transformed GP equation
\begin{eqnarray}
 \Mi\hbar\,\pdiff{\psi_\Mlabel\Lambda}{\tau}
 =& -\frac{\hbar^2}{2m}\,\left[\Lambda^{-\tp}(\tau) \bnabla_{\!\bzeta}\right]^2
   \psi_\Mlabel\Lambda(\tau,\bzeta) &
   \nonumber\\
 &+\, \frac{1}{\det \Lambda(\tau)}\left[\,
   \frac{m}{2}\, \bzeta^{\tp}\Omega^{2\!}(0)\, \bzeta
   + g\, |\psi_\Mlabel\Lambda(\tau,\bzeta)|^2
   - \mu\right]\psi_\Mlabel\Lambda(\tau,\bzeta)
\label{ScaledGP}
\end{eqnarray}
for the wave function $\psi_\Mlabel\Lambda(\tau,\bzeta)$, where we have made use of the shorthand notation $\Lambda^{-\tp} \equiv (\Lambda^{-1})^\tp$ and the requirement that the matrix $\Lambda(\tau)$ obeys
the nonlinear matrix differential equation
\begin{equation}
 \Lambda^\tp\!(\tau)
 \left[\diffz{\!\Lambda}{\tau}+\Omega^{2\!}(\tau)\,\Lambda(\tau)\right]
 = \frac{\Omega^{2\!}(0)}{\det \Lambda(\tau)} \,.
\label{MatrixDGL}
\end{equation}
The additional assumption that at $\rr=\bzeta$ and $t=0$ we have $\psi_\Mlabel D(0,\rr)=\psi_\Mlabel\Lambda(0,\rr)$ yields the initial conditions 
\begin{equation}
  \label{InitialMatrix}
  \Lambda(0)
  = \id\quad\mbox{and}\quad\left.
  \diff{\Lambda}{\tau}\right|_{\tau=0}=0
\end{equation}
for the nonlinear matrix differential equation \eqref{MatrixDGL}.
As discussed in~\ref{sec:app:irrotationality-condition}, the symmetry of the matrix $A(\tau)$, Eq. \eqref{Def_A}, is connected to the so-called irrotationality condition 
\begin{equation}
  \Lambda^\tp(\tau)\,\diff{\Lambda}{\tau}=\diff{\Lambda^\tp}{\tau}\, \Lambda(\tau)\,,
 \label{eq:IrrotCondition}
\end{equation}
that gives rise to $d(d-1)/2$ constants of motion of the matrix differential equation \eqref{MatrixDGL}. 

So far, we have made no approximation in deriving the affinely transformed GP equation~\eqref{ScaledGP}. 
Although it looks much more complicated than Eq. \eqref{GPequationD},
the solutions of Eq.~\eqref{ScaledGP} show almost no time dependence 
within the TF regime. This observation enables us to establish an efficient description of the BEC 
dynamics, as discussed in subsection \ref{TF_app}.

\subsubsection{The affinely transformed wave function}
By combining the two transformations given by Eqs. \eqref{DefpsiD}, \eqref{lineartrafo} and \eqref{DefpsiLambda}, 
we obtain the following relation between the original and the affinely transformed wave function 
\begin{equation}
  \label{eq:volleTrafo}
  \psi(t,\rr)
 = \frac{1}{\sqrt{\det\Lambda(t)}}\;
   \Me^{\Mi\Phi(t,\rr)}
   \psi_\Mlabel\Lambda\left(t,\Lambda^{-1}(t)\left[\rr-\calR(t)\right]\right)\,,
\end{equation}
where we have introduced the local phase
\begin{equation}
 \fl
  \Phi(t,\rr)\equiv \frac{1}{\hbar}\left\{
      \calS_1(t)-\beta(t)+\calP(t)\left[\rr-\tfrac{1}{2}\calR(t)\right]
      + \frac{m}{2}\left[\rr-\calR(t)\right]^{\tp}C(t)\left[\rr-\calR(t)\right]
    \right\}
\label{eq:RelPhase}
\end{equation}
and the time-dependent, symmetric matrix
\begin{equation}
  C(t)\equiv \diff{\Lambda}{t}\,\Lambda^{-1\!}(t).
  \label{Def_C}
\end{equation}
Since the matrix $C(t)$ determines the time evolution of 
the quadratic phase term, we refer to it as ``quadratic phase matrix''. For later purposes we also 
note, that by inverting Eq.~\eqref{eq:volleTrafo}, we find the affinely transformed wave function 
$\psi_\Mlabel\Lambda(t,\rr)$ in terms of $\psi(t,\rr)$ via
\begin{equation}
  \label{eq:volleInverseTrafo}
  \psi_\Mlabel\Lambda(t,\rr) = \sqrt{\det{\Lambda(t)}}\;\Me^{-\Mi\Phi(t,\Lambda(t)\rr+\calR(t))}
    \psi\left(t,\Lambda(t)\rr\!+\!\calR(t)\right)\,.
\end{equation}

In summary, the affine transformation of the macroscopic wave function is  
realized by Eq.~\eqref{eq:volleTrafo} and leads to the affinely transformed GP equation~\eqref{ScaledGP}.
In order to determine $\psi(t,\rr)$, one has to solve the time-dependent partial differential 
Eq.~\eqref{ScaledGP} together with the ordinary differential equations \eqref{eq:equationsOfMotion} 
and \eqref{MatrixDGL} for the center-of-mass 
variables $\calR(t)$, $\calP(t)$ and the adaptive matrix $\Lambda(t)$, respectively. 
Throughout this article, we refer to this generalization of the standard scaling method~\cite{kagan96,castin96,kagan97} 
as the ``affine approach''.

\subsubsection{Non-rotating trap: connection to the scaling approach}
We continue by pointing out the connection between our affine approach and the
scaling method introduced by \cite{kagan96,castin96,kagan97}. 
We consider a non-rotating trap and assume for simplicity that the principal axes of
the harmonic potential coincide with the coordinate axes. This can always be achieved by an appropriate 
choice of coordinates for which the trap matrix $\Omega^{2}(t)$ in Eq. \eqref{eq:potential} possesses 
the diagonal form
\begin{equation}
  \Omega^{2}(t)\equiv\op{diag}\left[\omega_1^{2}(t),\ldots,\omega_d^{2}(t)\right],
\end{equation}
with $\omega_i(t)$ being the trapping frequency along the $x_i$ direction.
Accordingly, we assume that the adaptive matrix $\Lambda(\tau)$ is of diagonal form and substitute
$\Lambda(\tau)\equiv\op{diag}\left[\lambda_1(\tau),\ldots,\lambda_d(\tau)\right]$ into Eq. \eqref{MatrixDGL} which 
yields the coupled nonlinear differential equations
\begin{equation}
 \label{eq:Castin-Kagan}
  \diffz{\lambda_i}{\tau} + \omega_i^2(\tau)\lambda_i(\tau)
  = \frac{\omega^2_i(0)}{\lambda_i(\tau)\,
    \prod_{k=1}^{d}\!\lambda_k(\tau)}\,,
\end{equation}
where $i\in 1,\dots,d$.
The functions $\lambda_i(\tau)$ characterize the time evolution of the condensate
in terms of an individual scaling along the three
principal axes of the potential. From Eq.~\eqref{InitialMatrix}, we obtain for the corresponding initial conditions
\begin{equation}
 \label{eq:Castin-Kagan-initial conditions}
  \lambda_i(0)
  = 1 \quad\mbox{and}\quad \diff{\lambda_i}{\tau}\Big|_{\tau=0} = 0\,.
\end{equation} 
Clearly, the irrotationality condition, Eq.~\eqref{eq:IrrotCondition}, is automatically satisfied 
for a diagonal adaptive matrix~$\Lambda(\tau)$. For the case $d=3$, Eqs. \eqref{eq:Castin-Kagan} and \eqref{eq:Castin-Kagan-initial conditions} precisely rephrase the well-known equations derived for the scaling factors in Ref. \cite{castin96,kagan97}.

\subsubsection{Interpretation of the classical trajectory\label{sec:rho}}

During the preparation of the BEC by laser and evaporative cooling at $t\leq 0$, we denote by $\brho(t)$
the position of the minimum of the full external potential $V(t,\rr)$ which acts on the atomic cloud. For this reason, the classical trajectory $\brho(t)$
is found as a solution of the equation
\begin{equation}
\label{eq:trap_minimum}
\bnabla_{\rr} V(t,\rr)|_{\rr=\brho(t)}=0
\end{equation}
for all times $t\leq 0$. Since the trapping potential $V(t,\rr)$ is located in an Earth-bound laboratory, a capsule freely falling in a drop tower or a satellite in space, inertial effects due to the local acceleration and rotation of the 
comoving frame of reference attached to the trapping potential $V(t,\rr)$ along $\brho(t)$ do occur in general. However, here we assume that these inertial effects can be neglected for all times $t\leq 0$ due to the dominating influence of the trapping potential and the repulsive interaction of the atoms.

As pointed out in section~\ref{sec:com_Elimination}, the time-dependent vectors $\calR(t)$ and $\calP(t)$ are associated 
with the center-of-mass motion of the atomic cloud. This interpretation implies that the initial conditions $\calR(0)$ and $\calP(0)$ for the classical equations of motion~\eqref{eq:equationsOfMotion} are directly linked to the preparation of the BEC in the external potential $V(t,\rr)$ according to
\begin{eqnarray}
 \label{eq:com_initialCond}
  \calR(0)&=&\brho(0)\,,\nonumber\\
  \calP(0)&=&m\dot{\brho}(0)\,.
\end{eqnarray}
In other words, the center of mass of the initial BEC at $t=0$ is supposed to be at rest within 
the comoving frame of reference attached to the external potential $V(t,\rr)$ at the position $\brho(0)$.

After the initial preparation phase, the dynamical evolution of the BEC for times $t>0$ can be analyzed based on two different 
interpretations of the classical trajectory $\brho(t)$. In case one is interested in the relative motion of the center of mass of the BEC with respect to the trap minimum, the trajectory $\brho(t)$ can still be associated with the minimum of the external potential $V(t,\rr)$ and Eq.~\eqref{eq:trap_minimum} has to be satisfied by $\brho(t)$ also for $t>0$.

In terms of a semiclassical approach, one can likewise associate $\brho(t)$ for all $t>0$ with the center-of-mass motion of the atomic cloud itself. In this case, the trajectory $\brho(t)$ is determined as 
the solution of the classical equation of motion 
\begin{equation}
\label{eq:semiclassical_eq_of_motion}
  m\ddot{\brho}(t) = - \bnabla_{\rr} V(t,\rr)|_{\rr=\brho(t)}
\end{equation}
for $t>0$ and describes the semiclassical motion of the center of mass of the BEC in the full external 
potential $V(t,\rr)$. Thus, the time-dependent vectors $\calR(t)$ and $\calP(t)$ follow directly from
$\brho(t)$ via
\begin{eqnarray}
 \label{eq:com_from_rho}
  \calR(t)&=&\brho(t)\,,\nonumber\\
  \calP(t)&=&m\dot{\brho}(t)\,,
\end{eqnarray}
whereas the classical equations of 
motion, Eq.~\eqref{eq:equationsOfMotion}, mathematically coincide with Eq.~\eqref{eq:semiclassical_eq_of_motion}.

\subsection{The macroscopic wave function in the time-dependent Thomas-Fermi approximation}
\label{TF_app}

We start this subsection by recalling the TF approximation
\cite{Fetter1998,pethick02} for the ground state of a BEC in a harmonic trap. 
We then apply this method also to the case of a time-dependent rotating trap in order to
provide an efficient description of the dynamics of the macroscopic wave
function of a BEC within the TF regime \cite{kagan96,castin96,kagan97,storey00,DGPS99}. 

In addition, we present a simple relationship between the spatial density distribution of a BEC in three
dimensions and its corresponding integrated density distributions in one and two
dimensions, which allows us to fully characterize the BEC dynamics within the TF regime by a sequence of three mutually orthogonal time-of-flight pictures. 
We conclude this subsection by briefly discussing an alternative description of the
internal dynamics of a rotating BEC within the TF regime~\cite{Edwards2002}.

\subsubsection{Thomas-Fermi approximation for the initial ground state}

Without loss of generality, we assume that $V(0,\brho(0)) = 0$ at the initial time $t=0$. 
Furthermore, we recall that $\bgg(0)=0$ for the harmonic 
potential~\eqref{eq:potential} due to the validity of Eq.~\eqref{eq:trap_minimum}. 
Taking also Eq.~\eqref{eq:com_initialCond} into account, the ground state $\phi(\rr)$ of a 
BEC at $t=0$ is defined as the solution of the stationary GP equation
\begin{equation}
  \label{eq:gpEquation}
    \hspace{-1cm}
\mu\, \phi(\rr) = \left\{-\frac{\hbar^2}{2m}\bnabla^2_{\rr}+
\frac{m}{2}[\rr-\calR(0)]^{\tp}\Omega^{2}(0)[\rr-\calR(0)]+ g|\phi(\rr)|^2\right\}\phi(\rr),
\end{equation}
with $\mu$ being the chemical potential of the ground state.

The TF regime is characterized  by a dominant contribution of the atomic interactions 
to the total energy of the BEC \cite{Fetter1998,pethick02}. In this case, 
the kinetic energy term in the stationary GP equation \eqref{eq:gpEquation} only accounts for a 
negligible contribution to the total energy in comparison with those 
arising from the harmonic potential and the atom-atom interactions. For this reason, within the TF approximation one simply neglects 
the kinetic energy term which gives rise to the approximate ground state of the BEC
\begin{equation}
  \label{eq:initialTFwavefunc}
  \phi_\Mlabel{TF}(\rr)= \frac{1}{\sqrt{g}}\left\{\mu_\Mlabel{TF}
    - \frac{m}{2}[\rr-\calR(0)]^{\tp}\Omega^{2\!}(0)[\rr-\calR(0)]\right\}^{\frac12}_+ \, \Me^{\frac{\Mi}{\hbar}\calP(0)\,\rr} \,.
\end{equation}
Here, the chemical potential $\mu_{\Mlabel{TF}}$ within the TF approximation 
\begin{equation}
  \label{eq:chemicalPotentialTF}
  \mu_\Mlabel{TF}
  = \frac{m}{2}\left(\frac{2\,\GammaF\left(2+\frac{d}{2}\right)}{\pi^{d/2}}
  \frac{Ng}{m}\sqrt{\det\Omega^{2}(0)}\right)^{\textstyle{\frac{2}{d + 2}}}
\end{equation}
is determined by the normalization condition~\eqref{eq:Normalization} for the wave function $\phi_\Mlabel{TF}(\rr)$, 
$\GammaF(x)$ denotes the Gamma function and 
\begin{equation}
 \{x\}^\alpha_{+}\equiv x^\alpha\Theta(x)
\end{equation}
stands for the positive part of $x$ to the power of $\alpha$, $\alpha>0$, with $\Theta(x)$ being the Heavyside function.
The additional phase $\Mi\calP(0)\,\rr/\hbar$ in Eq.~\eqref{eq:initialTFwavefunc} 
is due to the non-vanishing momentum $\calP(0)$, Eq.~\eqref{eq:com_initialCond}, of the initial BEC as discussed in section~\ref{sec:rho}.

The ground state within the TF approximation $\phi_\Mlabel{\Lambda,TF}(\bzeta)$ rewritten in terms of the adapted coordinates $\bzeta$ follows directly from Eqs. \eqref{eq:volleInverseTrafo} and \eqref{eq:initialTFwavefunc} 
according to
\begin{equation}
\fl
  \phi_\Mlabel{\Lambda,TF}(\bzeta)=\Me^{-\Mi\Phi(0,\bzeta+\calR(0))} \phi_\Mlabel{TF}\left(\bzeta+\calR(0)\right)
  = \frac{1}{\sqrt{g}}\left\{
    \mu_\Mlabel{TF}
    - \frac{m}{2}\,\bzeta^{\tp}\Omega^{2}(0)\,
      \bzeta\right\}^{\frac12}_+ \, \Me^{\frac{\Mi}{2\hbar}\calR(0)\calP(0)}.
  \label{eq:initialTFwavefuncLambda}
\end{equation}
Besides the initial conditions for the adaptive matrix $\Lambda(t)$, Eq. \eqref{InitialMatrix}, we have used the defining Eqs.~\eqref{eq:PhaseS} for $k=1$,~\eqref{Def_beta} and~\eqref{Def_C} to find $\Phi(0,\rr)=\calP(0)\left[\rr-\tfrac{1}{2}\calR(0)\right]/\hbar$ for the local phase~\eqref{eq:RelPhase} at $t=0$ in the derivation of Eq.~\eqref{eq:initialTFwavefuncLambda}.

\subsubsection{Time-dependent Thomas-Fermi approximation for the dynamical evolution}
\label{sec:effective_description}

Within the time-dependent TF approximation \cite{castin96,kagan97}, one neglects the kinetic energy term in the affinely transformed GP equation \eqref{ScaledGP} for all times $\tau\geq 0$ in order to arrive at the approximate, ordinary differential equation
\begin{equation}
 \Mi\hbar\,\pdiff{\psi_\Mlabel\Lambda}{\tau}
 \approx\, \frac{1}{\det \Lambda(\tau)}\left[
   \frac{m}{2} \,\bzeta^{\tp}\Omega^{2\!}(0)\, \bzeta
   + g\, |\psi_\Mlabel\Lambda(\tau,\bzeta)|^2
   - \mu\right]\psi_\Mlabel\Lambda(\tau,\bzeta) .
\label{ScaledGPTF}
\end{equation}
As we will show below, the solution $\psi_\Mlabel\Lambda(\tau,\bzeta)$ of Eq.~\eqref{ScaledGPTF} 
is time-independent and therefore agrees with the approximate ground state~$\phi_\Mlabel{\Lambda,TF}(\bzeta)$, Eq.~\eqref{eq:initialTFwavefuncLambda}, for all times $\tau \geq 0$
\begin{equation}
 \psi_\Mlabel\Lambda(\tau,\bzeta)=\psi_\Mlabel\Lambda(0,\bzeta)\approx\phi_\Mlabel{\Lambda,TF}(\bzeta)\,.
\label{eq:Notimeevol_for_trans_wavefct}
\end{equation}
This observation is the most important result of the time-dependent TF approximation and 
it manifests its full strength by providing an efficient and accurate description of
the BEC dynamics within the TF regime.

In order to prove the validity of Eq.~\eqref{eq:Notimeevol_for_trans_wavefct}, 
we first multiply the ordinary differential equation~\eqref{ScaledGPTF} 
by $\psi^*_\Mlabel\Lambda(\tau,\bzeta)$ and subtract 
the imaginary part of the resulting equation to find
\begin{equation}
 \pdiff{}{\tau}\Big(|\psi_\Mlabel\Lambda(\tau,\bzeta)|^2\Big)\approx 0\,.
\end{equation}
Thus, the absolute value of the affinely transformed wave function does not change
in time and  we conclude that
\begin{equation}
|\psi_\Mlabel\Lambda(\tau,\bzeta)|^2=|\psi_\Mlabel\Lambda(0,\bzeta)|^2\approx|\phi_\Mlabel{\Lambda,TF}(\bzeta)|^2\;\;\; {\rm and}\;\;\;
\mu\approx\mu_\Mlabel{TF}\,.
\label{eq:TimeDependentTFAssumptions}
\end{equation}
We infer from the Eqs.~\eqref{eq:initialTFwavefuncLambda} and \eqref{eq:TimeDependentTFAssumptions}, 
that the square bracket on the right hand side of the ordinary differential equation~\eqref{ScaledGPTF} vanishes, giving rise to $\partial\psi_\Mlabel\Lambda/\partial \tau=0$ which finally justifies the validity of Eq.~\eqref{eq:Notimeevol_for_trans_wavefct}.

To find the above mentioned efficient description of the BEC dynamics, 
we take advantage of Eq.~\eqref{eq:volleTrafo} with the function $\psi_\Mlabel\Lambda(\tau,\bzeta)$ being determined by 
combining the Eqs.~\eqref{eq:Notimeevol_for_trans_wavefct} and \eqref{eq:initialTFwavefuncLambda}. As a result, we obtain the central expression
\begin{equation}
\fl
  \label{eq:solutionTF}
  \psi(t,\rr) \approx  \psi_\Mlabel{TF}(t,\rr)=
  \frac{\Me^{\Mi\Phi_{\Mlabel{TF}}(t,\rr)}}{\sqrt{g\,\det\Lambda(t)}}\;
  \left\{\mu_\Mlabel{TF} - \frac{m}{2}
  \left[\rr-\calR(t)\right]^{\tp}\Sigma^{-1}(t)\left[\rr-\calR(t)\right]\right\}^{\frac12}_+\,,
\end{equation}
that characterizes the time evolution of a BEC within the time-dependent TF approximation. 
In Eq.~\eqref{eq:solutionTF} we have introduced the new local phase
\begin{equation}
 \hspace{-1cm}
  \Phi_{\Mlabel{TF}}(t,\rr)\equiv \frac{1}{\hbar}\left\{
      \calS_2(t)-\beta(t)+\calP(t)\,\rr 
      + \frac{m}{2}\left[\rr-\calR(t)\right]^{\tp}C(t)\left[\rr-\calR(t)\right]
    \right\}\,,
\label{eq:RelPhase_TF}
\end{equation}
where the generalized action $\calS_2(t)$ is given by Eq.~\eqref{eq:PhaseS} for $k=2$. Moreover, we have defined 
the so-called ``TF matrix'' 
\begin{equation}
  \label{eq:TF_Matrix}
  \Sigma(t)  = \Lambda(t)\,\left[\Omega^{2\!}(0)\right]^{-1}\Lambda^{\tp\!}(t),
\end{equation}
which is positive definite. 

According to Eq.~\eqref{eq:solutionTF}, the boundary at which the spatial density distribution of the BEC vanishes 
corresponds to an ellipse if $d=2$ or to an ellipsoid if $d=3$. The TF matrix~\eqref{eq:TF_Matrix} defines the orientation and semi-principal axes of this ellipse or ellipsoid whereby its points $\rr\in\mathds{R}^d$ satisfy the condition 
${[\rr-\calR(t)]^{\tp}\Sigma^{-1}(t)[\rr-\calR(t)]=2\mu_\Mlabel{TF}/m}$.
The eigenvalues $\sigma_i$ of the TF matrix $\Sigma(t)$ determine the lengths 
of the semi-principal axes, which coincide with the individual TF radii $r_i$ of the BEC, via $r_i=\sqrt{2\mu_\Mlabel{TF}\,\sigma_i/m}$.

\begin{figure}[h]
 \begin{center}
 \includegraphics[scale=1]{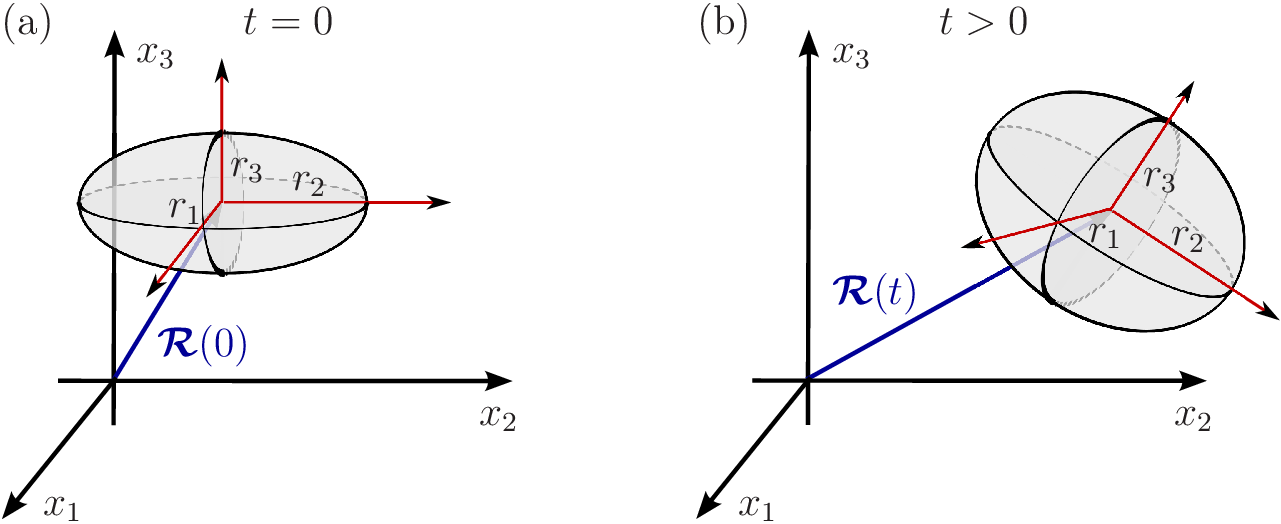}
 \caption{Boundaries of an ellipsoidal density distribution of a BEC 
 within the time-dependent TF approximation. At $t=0$ the BEC is centered around $\calR(0)$ and 
 the principal axes of its boundary are parallel to the coordinate axes (a). 
 Modifications in the quadratic trapping potential at $t > 0$ result in a center-of-mass motion $\calR(t)$ and 
 changes in the ellipsoidal shape and the TF radii $r_i$ of the BEC (b).}
 \label{Fig:BEC_Ellipsoid}
 \end{center}
 \end{figure}

Fig.~\ref{Fig:BEC_Ellipsoid} illustrates the time evolution of such an ellipsoid in three dimensions.
The initial density distribution of the ground state is centered around $\calR(0)$ with its principal axes being parallel 
to the coordinate axes, as shown in Fig. \ref{Fig:BEC_Ellipsoid}(a). 
Changes in the orientation and position of the quadratic potential~\eqref{eq:potential} give rise to a center-of-mass motion 
$\calR(t)$ and a transformation of the elliptic contour of the BEC density distribution via the time evolution of the TF matrix~$\Sigma(t)$, as depicted in Fig. \ref{Fig:BEC_Ellipsoid}(b).

We summarize this section by pointing out once more that the expression given by Eq. \eqref{eq:solutionTF} provides a valuable description of the time evolution of a  rotating BEC within the TF regime. Having the time-dependent TF approximation in mind, one might be tempted to believe that Eq. \eqref{eq:solutionTF} is only valid for times 
shortly after~$t=0$. However, Eq. \eqref{eq:solutionTF} yields very accurate results 
also for freely expanding BECs, since in this case the adaptive matrix $\Lambda(\tau)$
grows linearly in time, as discussed in section~\ref{sec:free-expansion-isotropic}, and therefore strongly reduces the influence of the kinetic energy term in Eq.~\eqref{ScaledGP}, the so-called quantum pressure.

\subsubsection{Integrated density distributions for time-of-flight pictures}
\label{sec:integratedDensity}

In practice, absorption imaging techniques \cite{Reinaudi2007} are the prevalent method to study the dynamics of a three-dimensional BEC. Identically prepared BECs are illuminated by a laser field at different times of flight
and their shadows are detected by a CCD camera resulting in a sequence of two-dimensional pictures of the spatial density
distribution of a BEC. For this reason, these techniques provide only partial information about the orientation and size of
the original three-dimensional density distribution.

In this subsection we discuss the one- and two-dimensional density distributions that result
from an originally three-dimensional density profile of a BEC within the TF regime. 
In addition, we sketch the determination of the underlying three-dimensional density distribution 
from three mutually orthogonal time-of-flight pictures of a single BEC. 
The high efficiency of our approach is due to the knowledge of the density profile of the macroscopic wave function, Eq.~\eqref{eq:solutionTF}. In case one does not have any {\it a priori} knowledge about 
the wave function under consideration, a reconstruction is still possible 
with the help of the Radon transformation~\cite{Gindikin1994,Schleich2001}.

We start with the density distribution of a three-dimensional BEC within the TF regime
\begin{equation}
|\psi_\Mlabel{TF}(t,\rr)|^2=
  \frac{1}{g\,\det\Lambda(t)}
  \left\{\mu_\Mlabel{TF}-\frac{m}{2}\left[\rr\!-\!\calR(t)\right]^{\tp}\Sigma^{-1}(t)\left[\rr\!-\!\calR(t)\right]\right\}_+\,,
 \label{eq:TimeEvolTF_Density}
\end{equation}
which directly follows from Eq.~\eqref{eq:solutionTF}.
The two-dimensional integrated density distribution in the $x$-$y$ plane 
\begin{equation}
  \label{eq:integratedDensityDefinition2d}
  n^\Mlabel {(2D)}_\Mlabel{TF}(t,x,y)\equiv \int_{\Reals} 
    \abs{\psi_\Mlabel{TF}(t,\rr)}^2\,\Mdiff z
\end{equation}
can be evaluated as discussed in~\ref{sec:appDensityIntegration} and reads
\begin{equation}
  \fl
  n^\Mlabel{(2D)}_\Mlabel{TF}(t,x,y)
  \!=\! \frac{4\sqrt{2}}{3g\sqrt{\vphantom{\dot\Sigma}m\det\bar\Sigma(t)\det\Omega^{2\!}(0)}}
    \left\{
      \mu_\Mlabel{TF}
      - \frac{m}{2}\left[\bar\rr \!-\! \bar{\calR}(t)\right]^{\!\tp}
        \!\bar{\Sigma}^{-1}(t) \!\left[\bar\rr \!-\! \bar{\calR}(t)\!\right]
    \right\}_+^{\frac32}\!\,.
 \label{eq:integratedDensit2d}
\end{equation}
Here, we have introduced the reduced vectors~$\bar\rr\equiv(x,y)^\tp$ and $\bar{\calR}\equiv(\mathcal{R}_1,\mathcal{R}_2)^\tp$ together with the symmetric submatrix
\begin{equation}
\bar\Sigma\equiv\left(
\begin{array}{ll}
\Sigma_{11} & \Sigma_{12}\\
\Sigma_{12} & \Sigma_{22}
\end{array}
\right)
\in
\mathds{R}^{2\times 2}
\label{eq:SubSigma_z}
\end{equation}
which is obtained from the TF matrix \(\Sigma\), Eq. \eqref{eq:TF_Matrix}, 
by removing the row~$\Sigma_{3i}$ and the column~$\Sigma_{i3}$ that correspond
to the $z$-coordinate. In a similar way, the two-dimensional integrated density 
distributions $n^\Mlabel {2D}_\Mlabel{TF}(t,y,z)$ and $n^\Mlabel {2D}_\Mlabel{TF}(t,x,z)$
depend on the two-dimensional submatrices in which the rows and columns
corresponding to the $x$- and $y$-coordinate have been eliminated, respectively.

For the one-dimensional integrated density distribution along the $x$-axis
\begin{equation*}
  n^\Mlabel {(1D)}_\Mlabel{TF}(t,x)
  = \int_{\Reals}\int_{\Reals} 
    \abs{\psi_\Mlabel{TF}(t,\rr)}^2 \,\Mdiff y\,\Mdiff z
\end{equation*}
we obtain
\begin{equation}
  n^\Mlabel {(1D)}_\Mlabel{TF}(t,x)
  = \frac{\pi}{m g\,\sqrt{\vphantom{\dot\Sigma}\Sigma_{11}(t)\det\Omega^{2\!}(0)}}
    \left\{
      \mu_\Mlabel{TF}
      - \frac{m}{2 \Sigma_{11}(t)} 
        \left[x \!-\! \mathcal{R}_1(t)\right]^{\!2} \right\}_+^2\,.
\label{eq:integratedDensit1d}
\end{equation}
The one-dimensional integrated density distributions 
$n^\Mlabel {(1D)}_\Mlabel{TF}(t,y)$ and $n^\Mlabel {(1D)}_\Mlabel{TF}(t,z)$ possess a similar structure.
A general derivation of the previous results for arbitrary dimensions is presented in \ref{sec:appDensityIntegration}.

To reconstruct the unknown three-dimensional density distribution 
of a BEC within the TF regime, Eq. \eqref{eq:TimeEvolTF_Density}, from three mutually orthogonal time-of-flight pictures, 
one needs to fit the two-dimensional integrated density distributions 
$n^\Mlabel {(2D)}_\Mlabel{TF}(t,x,y)$, $n^\Mlabel {(2D)}_\Mlabel{TF}(t,y,z)$ and 
$n^\Mlabel {(2D)}_\Mlabel{TF}(t,x,z)$ to the corresponding experimental data and thereby determine 
the submatrix \eqref{eq:SubSigma_z} and its two other counterparts. 
These three submatrices provide us with all elements of the TF matrix $\Sigma$, 
where each diagonal element of $\Sigma$ is found twice. This redundancy can be used to check the quality 
of the data obtained from the three mutually orthogonal time-of-flight pictures.

\subsubsection{Alternative description of rotating and vortex-free condensates}

Our description of the internal dynamics of a BEC within the TF regime rests upon
the solution $\Lambda(t)$ of the nonlinear matrix differential equation \eqref{MatrixDGL} 
with the corresponding initial conditions, Eq. \eqref{InitialMatrix}. Based on the adaptive matrix $\Lambda(t)$, we  
then determine the TF matrix $\Sigma(t)$ and the quadratic phase matrix $C(t)$
defined by Eqs. \eqref{eq:TF_Matrix} and \eqref{Def_C}, respectively. The two symmetric matrices
$\Sigma(t)$ and $C(t)$ characterize the time evolution of the absolute value and the phase of the 
macroscopic wave function $\psi(t,\rr)$, Eq. \eqref{eq:solutionTF}. Whereas the TF matrix~$\Sigma(t)$ plays a crucial role 
in the context of time-of-flight pictures, the quadratic phase matrix~$C(t)$ is of great importance for 
BEC based interferometry \cite{Torii2000,Simsarian2000,Debs2011,Altin2013,Muentinga2013}.

One might wonder whether there is a way to study the BEC dynamics directly in terms of 
the experimentally more accessible matrices $\Sigma^{-1}(t)$ and $C(t)$ without a link to the adaptive matrix $\Lambda(t)$ at all. In fact, such an alternative description is possible and involves the solution of the two coupled first order nonlinear matrix differential equations
\begin{eqnarray}
\diff{\Sigma^{-1}}{t} = -C(t)\,\Sigma^{-1}(t) - \Sigma^{-1}(t)\,C(t) \,,
\label{eq:Diff_Sigma}\\
\diff{C}{t} = - C^{2\!}(t) - \Omega^{2\!}(t)+\Sigma^{-1}(t)\sqrt{\frac{\det{\Sigma^{-1}(t)}}{\det{\Omega^{2\!}(0)}}}
\label{eq:Diff_C}
\end{eqnarray}
with the corresponding initial conditions
\begin{eqnarray}
 \Sigma^{-1}(0)& = &\Omega^{2\!}(0)\,, \nonumber \\
 C(0)& = & 0.
 \label{eq:InitialCond_Sigma_C}
\end{eqnarray}
These first order differential equations can be derived by taking the time derivative on both sides of Eqs.~\eqref{Def_C} and~\eqref{eq:TF_Matrix} and using the matrix differential equation~\eqref{MatrixDGL}, 
the two identities $(d\Sigma^{-1}/dt)=-\Sigma^{-1} (d\Sigma/dt)\Sigma^{-1}$ and
${\det{\Lambda(t)}=\sqrt{\det{\Omega^{2\!}(0)}/\det{\Sigma^{-1}(t)}}}$ as well as the symmetry of the matrix $C(t)$. 

At present, only a simplified version \cite{Edwards2002} of the system of equations 
\eqref{eq:Diff_Sigma} and \eqref{eq:Diff_C} has been used to study the dynamics of a freely expanding BEC that was initially prepared in a rotating anisotropic harmonic trap~\cite{Hechenblaikner2002}. The trap was instantly turned off at the time $t=T_\Mlabel{off}$. 
In fact, by inserting the trap matrix $\Omega^{2\!}(t)=0$ for all $t\geq T_\Mlabel{off}$ into both Eqs. \eqref{eq:Diff_Sigma} and \eqref{eq:Diff_C}, 
it can be shown that the resulting matrix differential equations are 
mathematically equivalent to the equations of motion (12) and (13) derived in Ref. \cite{Edwards2002}. 
However, the approach presented in this article facilitates an analysis of the time evolution of BECs within the TF regime that goes far beyond the scope considered in Refs.~\cite{Edwards2002,Hechenblaikner2002}. 

Next, we briefly outline how to obtain $\Lambda(t)$ from the quadratic phase matrix $C(t)$ and the TF matrix $\Sigma(t)$. When the time evolution of~$C(t)$ is known, the adaptive matrix \(\Lambda(t)\) follows from Eq.~\eqref{Def_C} by simple integration of the first order matrix differential equation
\begin{equation*}
  \diff{\Lambda}{t}
  = C(t) \Lambda(t)
\end{equation*}
with the initial condition \(\Lambda(0)=\id\). In contrast, the knowledge of the TF matrix $\Sigma(t)$ does not 
suffice to fully determine the adaptive matrix \(\Lambda(t)\). From Eq.~\eqref{eq:TF_Matrix}, the latter can only be identified up to an arbitrary orthogonal matrix $U(t)$ via the relation
\begin{equation*}
  \Lambda(t)
  = \Sigma^{\frac{1}{2}}(t)\,U(t)\,\Omega(0) \,,
\end{equation*}
where the square root of $\Sigma(t)$ is defined in terms of its spectral decomposition. The specification of the orthogonal 
matrix $U(t)$ involves again the quadratic phase matrix~$C(t)$.

There are two main reasons that make our approach based on the adaptive matrix~$\Lambda(t)$ more suitable for the 
characterization of BEC dynamics within the TF regime than the alternative description just presented. First, the second order matrix differential equation \eqref{MatrixDGL} for $\Lambda(t)$ can be recast in the form of Hamilton's equations, as shown in subsection \ref{sec:Hamilton:Formalism}. 
This fact allows us to apply the full mathematical machinery available for Hamiltonian mechanics to study the time evolution of $\Lambda(t)$. Second, for studying the density distribution of a BEC near its surface at the scale of the corresponding healing length, 
the TF ground state given by Eq. \eqref{eq:initialTFwavefunc} provides no longer an adequate solution of Eq. \eqref{eq:gpEquation} and the macroscopic wave function Eq. \eqref{eq:solutionTF} does not suffice for this purpose. 
Hence, the TF matrix $\Sigma(t)$, which rests upon the validity of Eq. \eqref{eq:solutionTF}, 
loses its immediate physical significance. 
In contrast, the affine approach based on the adaptive matrix $\Lambda(t)$ 
provides a valuable tool to study scenarios that are beyond the scope of the TF approximation with the help of dedicated numerical simulations, as discussed in the next section.

\section{Application of the affine approach to numerical simulations}
\label{Sec:Numerical}
In this section we show how numerical simulations of the BEC dynamics can benefit from the affine approach and we likewise apply this approach to quantify the accuracy of the time-dependent TF approximation for several different scenarios. For simplicity, we present here the study of the dynamics of rotating BECs on a 2D grid.  We emphasize that full 3D simulations of BECs in the quasi-2D regime \cite{petrov-2000,salasnich-2002,mateo-2008} have also been performed, which verify the validity of our 2D results.

\subsection{Efficient simulation of the time evolution of a BEC}
Solving numerically the time-dependent GP equation \eqref{GPequation} in the case of a freely expanding BEC in the original coordinates $(t,\rr)$, one immediately faces the problem that the size of the condensate grows by several orders of magnitude after switching off the trap. To deal with this problem, a large and well resolved grid is required, leading to a huge increase of the computational costs of the numerical simulations, especially for two or three dimensional cases. 

Here we present an alternative approach to overcome this problem. Namely we are solving the affinely transformed GP equation \eqref{ScaledGP} in the adapted coordinates $(\tau,\bzeta)$, rather than the GP equation \eqref{GPequation} in the original coordinates $(t,\rr)$. Since the external and internal dynamics of the BEC are handled by the time dependence of the center-of-mass position $\calR(t)$ and the adaptive matrix $\Lambda(\tau)$, respectively, only the time evolution of the affinely transformed wave function $\psi_\Mlabel\Lambda(\tau,\bzeta)$ is left and can be computed very efficiently \cite{PhD_ME}. As a result, in order to obtain the macroscopic wave function $\psi(t,\rr)$ in the original coordinates it is sufficient to apply the transformation Eq. \eqref{eq:volleTrafo} once at the end of the simulation. Moreover, our method is not limited to the case of free expansion, but can enhance the efficiency of numerical simulations for various experimental scenarios such as rotating traps or delta-kick 
collimation \cite{ammann-1997}.

\subsection{Quantifying the accuracy of the time-dependent Thomas-Fermi approximation}
Since the time-dependent TF approximation plays a key role in deriving the analytic expression, Eq. \eqref{eq:solutionTF}, for the wave function $\psi(t,\rr)$ of a BEC, it is worthwhile to have a closer look at its accuracy. The time-dependent TF approximation is based on assumption~\eqref{eq:Notimeevol_for_trans_wavefct} meaning that the wave function in adapted coordinates $\psi_\Mlabel\Lambda(\tau,\bzeta)$ remains approximately in its initial value $\psi_\Mlabel\Lambda(0,\bzeta)$ and does not undergo any time evolution at all. With our numerical simulations we have quantified the validity of this assumption for two different scenarios involving rotating traps. 

We take advantage of the Bures distance to quantify how much the numerically obtained state $\psi_\Mlabel{\Lambda}(\tau,\bzeta)$ differs from the initial state $\psi_\Mlabel{\Lambda}(0,\bzeta)$. In the following, we first recall the definition of the Bures distance and its properties and then describe the 2D model of a rotating BEC. Finally, we discuss the results of the simulations and characterize the parameter regime where it is safe to apply the TF approximation.

\subsubsection{Bures distance}
The Bures distance \cite{Bures1969} of two macroscopic wave functions $\psi_1$ and $\psi_2$ is given by
\begin{equation}
 \label{B-definition}
  B(\psi_1,\psi_2)
  \das \left(2-\frac{2\abs{\braket{\psi_1}{\psi_2}}}{\sqrt{\braket{\psi_1}{\psi_1}\braket{\psi_2}{\psi_2}}}\right)^{\frac{1}{2}}=
  \left(2-\frac{2}{N}\abs{\braket{\psi_1}{\psi_2}}\right)^{\frac{1}{2}}\;,
\end{equation} 
where we have used the normalized condition given by Eq. \eqref{eq:Normalization}. First we note, that one obtains $B(\psi_1,\psi_2) = 0$ if $\psi_1 = \psi_2$ and $B(\psi_1,\psi_2) = \sqrt{2}$ if $\psi_1$ and $\psi_2$ are orthogonal. Second, for two $d$-dimensional spherically symmetric Gaussian wave packets 
$$
\psi_{\Mlabel{G}}(\bx,\sigma_{1,2}) =\frac{\sqrt{N}}{(2\pi\sigma_{1,2}^2)^{d/4}}\;\Me^{-\frac{\bx^2}{4\sigma_{1,2}^2}}
$$
with different widths $\sigma_1$ and $\sigma_2$, the Bures distance Eq. \eqref{B-definition} reads 
\begin{equation}
 \label{B-Gaussian}
 B_G^{(d)}\left(\psi_{\Mlabel{G}}(\bx,\sigma_1),\psi_{\Mlabel{G}}(\bx,\sigma_2)\right) = 
 \sqrt{2\left[1-\left(\frac{2\sigma_1\sigma_2}{\sigma_1^2+\sigma_2^2}\right)^{d/2}\right]}\,.
\end{equation}

In the case of two-dimensional Gaussian wave packets with the relative difference $\delta_{\sigma} \equiv (\sigma_2 - \sigma_1)/\sigma_1$ in their widths, Eq. \eqref{B-Gaussian} gives rise to
\begin{equation}
 B_\Mlabel{G}^{(2)}(\delta_\sigma) \equiv 
 B_G^{(2)}\left(\psi_{\Mlabel{G}}(\bx,\sigma_1),\psi_{\Mlabel{G}}(\bx,(1+\delta_\sigma)\sigma_1)\right) = 
 \frac{\delta_\sigma}{\sqrt{1 + \delta_\sigma + \frac{1}{2}\delta_{\sigma}^2}}\,. 
 \label{eq:Bures2DGaussians} 
\end{equation}
Thus, a small relative change $\delta_\sigma$ in the width of a two dimensional Gaussian wave packet results in a small Bures distance, that is $B_\Mlabel{G}^{(2)}(\delta_\sigma)\cong \delta_\sigma \ll 1$. Despite the fact that the macroscopic wave function of a BEC is not typically Gaussian, this estimate can help us to evaluate the time dependence of the remaining inner dynamics which is not included in the affine approach.

\subsubsection{2D model of a Bose-Einstein condensate in a rotating trap}
The numerical simulations discussed in this section have all been performed with a 2D model based on realistic values that are accessible in state-of-the-art experiments. Our system consists of atoms of mass $m$ which are condensed and harmonically trapped along the $x$- and $y$-direction with the frequencies $\omega_x$ and $\omega_y$, respectively. The time evolution of this system is described by the affinely transformed GP equation \eqref{ScaledGP} with $d=2$. The number of discretization steps in time and the physical length of the grid are chosen as a trade-off between accuracy and computational time.

The free parameters of our system are $(i)$ the anisotropy factor $\epsilon\equiv \omega_y/\omega_x$, $(ii)$ the strength $\bar{g} N$ of the zero-range interaction between the atoms, and $(iii)$ the final rotation rate $\dot{\varphi}_\Mlabel{end}$ of the trap. All quantities are measured with respect to the chosen time scale $1/\omega_x$ and the length scale $a_x \equiv\sqrt{\hbar/(m \omega_x)}$, respectively. Table \ref{tab:parameter} displays the parameters for the numerical simulations as well as the relations between the physical quantities and the free parameters. For convenience we refer to the dimensionless quantity $\bar{g}N = g_\Mlabel{2D} N /(\hbar \omega_x a_x^2)$ as the interaction strength as it combines the 2D coupling constant $g_\Mlabel{2D}$ with the number of particles $N$.

\begin{table}[h]
\begin{center}
\caption{List of the grid settings, physical trap frequencies and 2D-interaction strength in terms of the simulation parameters.}
\begin{tabular}{l l}
\\ \hline\hline \\
number of discretization steps & $N_{x}=N_{y}=2^7$ \\
physical length of the grid & $L_{x}=L_{y}=20 \,a_x$ \\
\\
trap frequency along the $x$-axis & $\omega_{x}$ \\
trap frequency along the $y$-axis & $\omega_{y} = \epsilon \,\omega_x$ \\
2D-interaction strength & $g_\Mlabel{2D} N = \bar{g}N \,\hbar \omega_x a_x^2 $ \\
\\ \hline\hline 
\end{tabular}
\label{tab:parameter}
\end{center}
\end{table}

It is worth emphasizing that our 2D model can be experimentally realized as a 3D disk-shaped BEC in a highly anisotropic trap \cite{Goerlitz2001,burger-2001,rychtarik-2004}, with the confinement length $a_z \equiv\sqrt{\hbar/(m \omega_z)}$ along the $z$-direction being much smaller than those along the $x$- and $y$-direction \cite{petrov-2000,salasnich-2002,mateo-2008}, where $\omega_{z}$ is the trap frequency along the $z$-axis. As a result, the 2D coupling constant $g_\Mlabel{2D}$ is determined by the 3D one $g_\Mlabel{3D}$ as $g_\Mlabel{2D} = g_\Mlabel{3D}/(\sqrt{2\pi}a_z)$, where $g_\Mlabel{3D} = 4\pi\hbar^2 a_\Mlabel{s}/m$ and $a_\Mlabel{s}$ is the $s$-wave scattering length. Using these relations, the results based on the full 3D simulations have been shown to be in good agreement with the results of the 2D simulations.

\subsubsection{Time evolution within a rotating trap}
\label{sec:num:rotation}
The starting point for all simulations is the ground-state wave function $\psi_\Mlabel{\Lambda}(0,\bzeta)$ of the BEC obtained by imaginary time propagation of Eq. \eqref{ScaledGP} with $\Lambda(\tau)=\mathds{1}$ in combination with the Newton method. For the time evolution itself, the affinely transformed GP equation \eqref{ScaledGP} is solved with an implicit Adams-Bashforth-Moulton multi-step algorithm \cite{book-numerical-recipes-2007}.

The first scenario in which the accuracy of the time-dependent TF approximation is checked numerically is the rotating trap. In this case the time-dependent trap matrix ${\Omega^2(\tau) = O(\tau) D \, O^\tp(\tau)}$ contains the diagonal matrix 
\begin{equation*}
 D = \left(\begin{array}{*{2}{c}}
    \omega_x^2 & 0 \\ 0 &  \epsilon^2 \omega_x^2 \\
  \end{array}\right)
\end{equation*}
with the time-independent trap frequencies and an orthogonal matrix 
\begin{equation*}
 O(\tau)
  = \left(\begin{array}{*{2}{c}}
    \cos{\varphi(\tau)} & \sin{\varphi(\tau)}  \\
    -\sin{\varphi(\tau)} & \cos{\varphi(\tau)}  \\
  \end{array}\right)
\end{equation*}
describing a rotation around the $z$-axis with the time-dependent angle $\varphi(\tau)$. 

\begin{figure}[h]
\begin{center}
\includegraphics[scale=1]{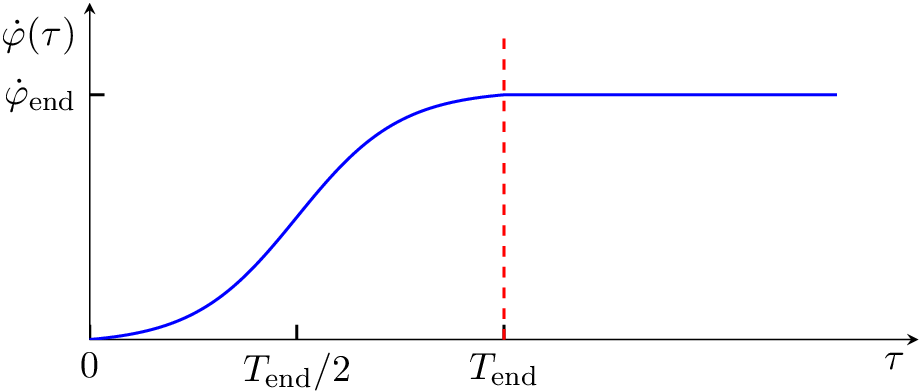}
\caption{The dependence of the rotation rate $\dot{\varphi}(\tau)$ of the harmonic trap on time $\tau$. 
After a smooth ramp within the characteristic time $T_\Mlabel{end}$, the rotation rate $\dot{\varphi}(\tau)$ reaches its maximum value $\dot{\varphi}_\Mlabel{end}$ and is
kept constant for all $\tau\geq T_\Mlabel{end}$.}
\label{fig:omega}
\end{center}
\end{figure}

Triggered by experiments \cite{Hechenblaikner2002,Muentinga2013}, the rotation rate $\dot{\varphi}(\tau)$ of the trap has been chosen to have a sigmoidal ramp for $0\leq \tau\leq T_\Mlabel{end}$, as depicted in Fig. \ref{fig:omega}, and to keep equal to the rotation rate $\dot{\varphi}_\Mlabel{end}$ for $\tau\geq T_\Mlabel{end}$. We assume that the characteristic time $T_\Mlabel{end}$ is large enough, that is $\omega_x T_\Mlabel{end}\gg 1$, to have an adiabatically slow increase of the rotation rate in order to avoid collective excitations of the BEC. In addition, the final rotation rate $\dot{\varphi}_\Mlabel{end}$ is kept significantly smaller than the trap frequency $\omega_x$ in order to avoid the generation of vortices. For the rest of this section we introduce the dimensionless characteristic time $\bar{T}_\Mlabel{end} \equiv \omega_x T_\Mlabel{end}/(2\pi)$ and use $\bar{T}_\Mlabel{end} = 15$ for all numerical simulations. 

\begin{figure}[h]
\begin{center}
\includegraphics[scale=1]{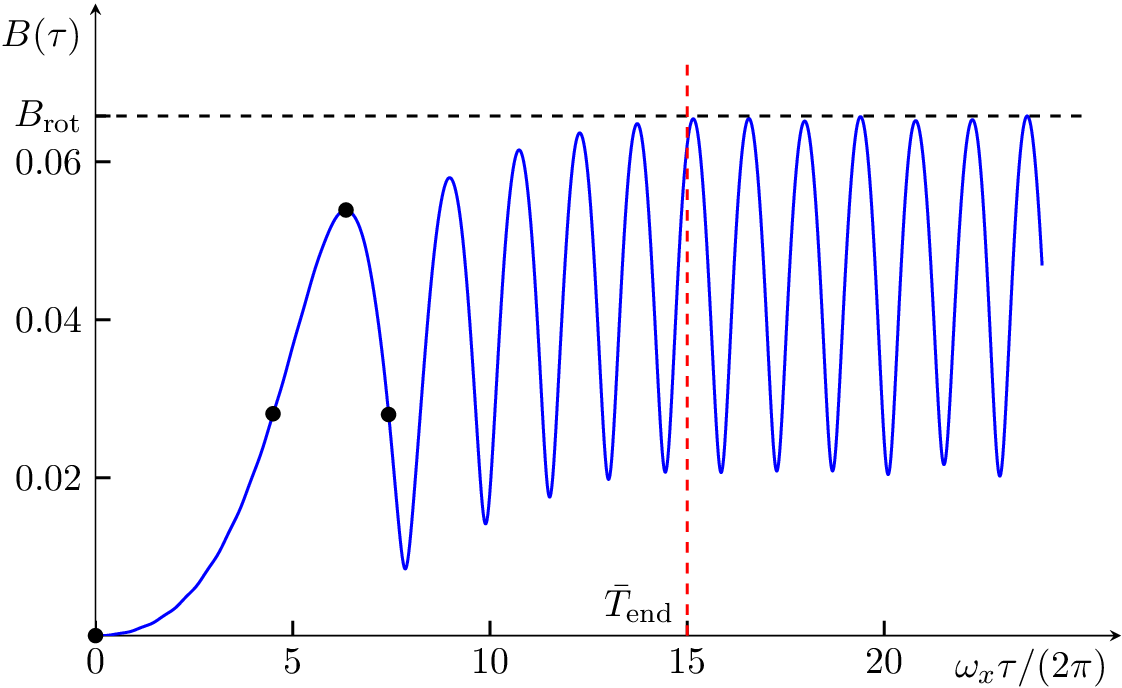}
\caption{Time-dependence of the Bures distance 
$B(\tau) = B\left(\psi_\Mlabel{\Lambda}(\tau,\bzeta),\psi_\Mlabel{\Lambda}(0,\bzeta)\right)$, Eq. \eqref{B-definition}, 
between the time evolved state $\psi_\Mlabel{\Lambda}(\tau,\bzeta)$ and the initial state $\psi_\Mlabel{\Lambda}(0,\bzeta)$ 
in adapted coordinates for a rotating trap with the anisotropy factor $\epsilon = 1.5$, 
the final rotation rate $\dot{\varphi}_\Mlabel{end} = 0.4\,\omega_x$ and the interaction strength $\bar{g}N = 100$.
The Bures distance $B(\tau)$ exhibits oscillations during the entire simulation, 
while the maxima of this oscillation grow until the rotation rate $\dot{\varphi}(\tau)$, 
shown in Fig. \ref{fig:omega}, reaches its maximum value $\dot{\varphi}_\Mlabel{end}$ 
at the characteristic time $\bar{T}_\Mlabel{end} = 15$, after which they remain constant at the value $B_\Mlabel{rot}$. 
The black circles indicate the times for which the corresponding density distributions are displayed 
in Fig. \ref{fig:density-rotation}.}
\label{fig:bures}
\end{center}
\end{figure}

Fig. \ref{fig:bures} shows the time dependence of the Bures distance $B(\tau) = B\left(\psi_\Mlabel{\Lambda}(\tau,\bzeta),\psi_\Mlabel{\Lambda}(0,\bzeta)\right)$, Eq. \eqref{B-definition}, between the time evolved state $\psi_\Mlabel{\Lambda}(\tau,\bzeta)$, being the numerical solution of Eq. \eqref{ScaledGP}, and the initial state $\psi_\Mlabel{\Lambda}(0,\bzeta)$ in adapted coordinates. The Bures distance $B(\tau)$ exhibits oscillations over the whole simulation time $\tau$ and the local maxima of this oscillation grow as long as the rotation rate increases until they reach their maximal value $B_\Mlabel{rot}$ at the characteristic time $\bar{T}_\Mlabel{end}$ and stay constant from that point on. Since the magnitude of the Bures distance is very small, $B_\Mlabel{rot}< 0.07$, the time evolved state does not differ significantly from the initial state.

For the parameters used in Fig. \ref{fig:bures} the Bures distance oscillates with the frequency $\omega_\Mlabel{Bures} = 1.8 \,\dot{\varphi}_\Mlabel{end}$ for $\tau>T_\Mlabel{end}$. Since the Bures distance, Eq. \eqref{B-definition}, only measures the absolute value of the overlap between two wave functions, the time evolved state $\psi_\Mlabel{\Lambda}(\tau,\bzeta)$ actually undergoes collective oscillations with the frequency $0.5\,\omega_\Mlabel{Bures}$. In general, the frequency of the collective oscillations caused by the rotation of the trap depends on the rotation rate $\dot{\varphi}(\tau)$ as well as the anisotropy factor $\epsilon$ of the trap. 

\begin{figure}[h]
 \begin{center}
 \includegraphics[scale=1]{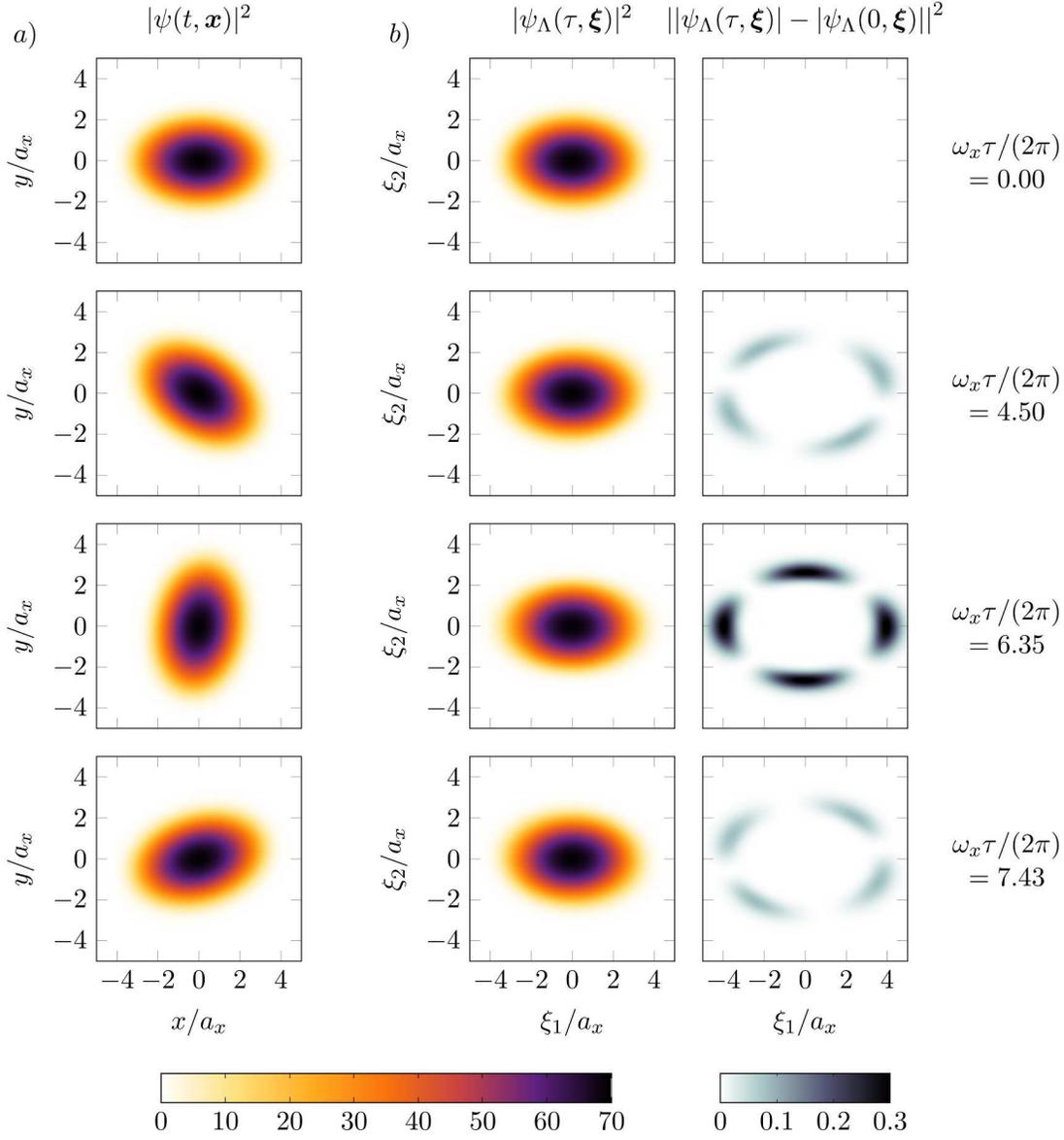}
 \caption{Density distributions $\abs{\psi(t,\rr)}^2$ and $\abs{\psi_\Mlabel{\Lambda}(\tau,\bzeta)}^2$ 
 as well as the difference $\abs{\abs{\psi_\Mlabel{\Lambda}(\tau,\bzeta)} - \abs{\psi_\Mlabel{\Lambda}(0,\bzeta)}}^2$ 
 of a rotating BEC plotted for different times as indicated in Fig. \ref{fig:bures}. 
 In original coordinates, Fig. \ref{fig:density-rotation} a), the density distribution undergoes a clockwise rotation, 
 while it stays almost constant in adapted coordinates, Fig. \ref{fig:density-rotation} b). 
 The residual dynamics in adapted coordinates are made visible in the third column by considering 
 the difference $\abs{\abs{\psi_\Mlabel{\Lambda}(\tau,\bzeta)} - \abs{\psi_\Mlabel{\Lambda}(0,\bzeta)}}^2$ 
 between the time evolved state $\psi_\Mlabel{\Lambda}(\tau,\bzeta)$ and the initial state $\psi_\Mlabel{\Lambda}(0,\bzeta)$. 
 Only at the edge of the condensate this difference obtains non-zero values while its magnitude corresponds 
 well with the oscillations of the Bures distance $B(\tau)$ shown in Fig. \ref{fig:bures}. 
 Movie 1 displaying the whole time evolution is available in the online supplementary material.}
 \label{fig:density-rotation}
  \end{center}
 \end{figure}

In Fig. \ref{fig:density-rotation} we contrast the dynamics of a rotating BEC in original and adapted coordinates while the whole time evolution is shown in movie 1, which is available in the online supplementary material. 
Fig. \ref{fig:density-rotation}a depicts the two-dimensional (non-integrated) density distributions $|\psi(t,\rr)|^2$
of a BEC in the original coordinates at different evolution times, whereas the first column of Fig. \ref{fig:density-rotation}b displays 
the corresponding density distributions $|\psi_\Mlabel{\Lambda}(\tau,\bzeta)|^2$ in the adapted coordinates. 
In contrast to the density distribution $|\psi(t,\rr)|^2$, which follows the clockwise rotation 
induced by the rotating trap, the density distribution $|\psi_\Mlabel{\Lambda}(\tau,\bzeta)|^2$ 
in adapted coordinates does not show any visible changes. 
To highlight the residual dynamics, we present in the second column of Fig. \ref{fig:density-rotation}b
the difference $||\psi_\Mlabel{\Lambda}(\tau,\bzeta)|-|\psi_\Mlabel{\Lambda}(0,\bzeta)||^2$ 
of the absolute values of the wave functions, which depends only on local effects of the density distributions and not 
on their phases. 
This difference has non-zero values only at the edge of the BEC 
forming a ring-like structure, which rotates clockwise in accordance with the density distribution $|\psi(t,\rr)|^2$ 
presented in Fig. \ref{fig:density-rotation}a. However, the magnitude of this residual dynamics is very small 
compared to the magnitude of the density distribution $|\psi_\Mlabel{\Lambda}(\tau,\bzeta)|^2$ and 
it oscillates in accordance with the Bures distance shown in Fig. \ref{fig:bures}. 

In summary, we have proven that the time-dependent TF approximation 
applied to the affinely transformed GP equation \eqref{ScaledGP} provides an accurate description of 
the dynamics of a rotating BEC. 
Only at the very edge of the BEC a minor amount of residual dynamics occurs that goes beyond the time-dependent TF approximation.
In subsection \ref{sec:num:dependence-sim-parameters} we study 
how the magnitude of this residual dynamics depends on the different setup parameters.

\subsubsection{Free expansion after switching off a rotating trap}
\label{sec:num:free-expansion}
As the second scenario to verify the time-dependent TF approximation, 
we have studied the subsequent free expansion of an initially rotating BEC.
Here, the setup is the same as in the first scenario 
with the exception that the trap is switched off at $\tau=T_\Mlabel{off}> T_\Mlabel{end}$, that is shortly after the final rotation rate $\dot{\varphi}_\Mlabel{end}$ is reached. Since the timing of the switch off has a strong influence on the final value $B_\Mlabel{free}$ of the Bures distance, we have varied $T_\Mlabel{off}$ within the first period of the Bures distance after $T_\Mlabel{end}$ for each set of parameters to find the maximal value of $B_\Mlabel{free}$.

\begin{figure}[h]
\begin{center}
\includegraphics[scale=1]{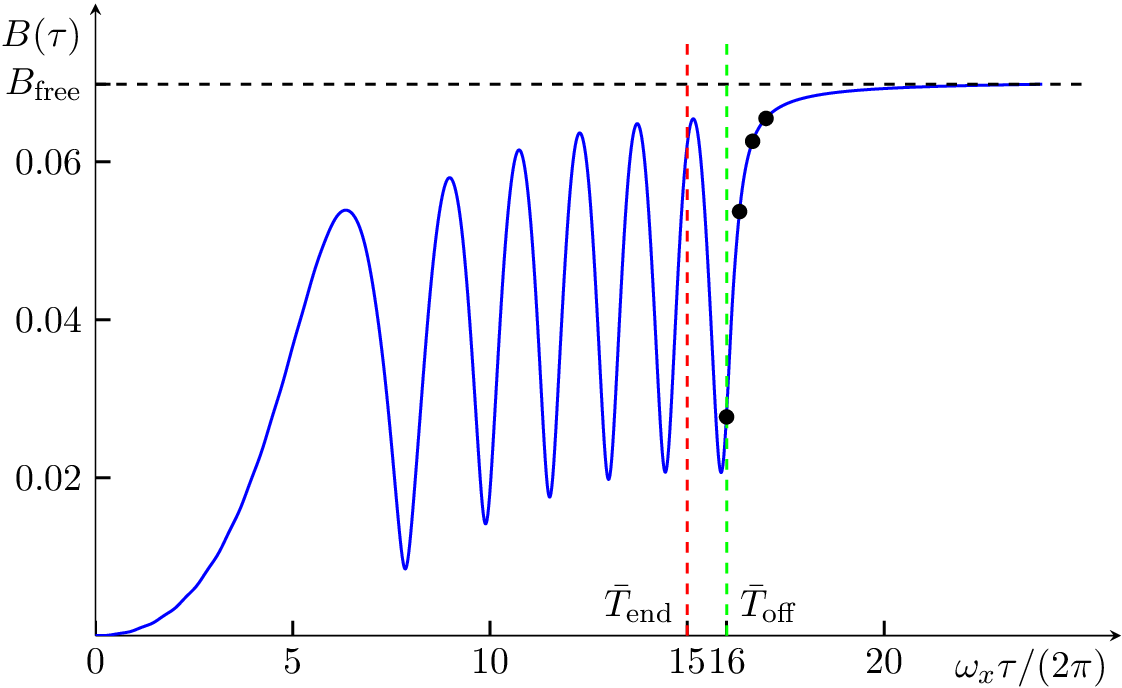}
\caption{Bures distance $B(\tau) = B(\psi_\Mlabel{\Lambda}(\tau,\bzeta),\psi_\Mlabel{\Lambda}(0,\bzeta))$ 
between the time evolved state and the initial state in adapted coordinates for free expansion 
after switching off a rotating trap with the anisotropy factor $\epsilon = 1.5$, the final rotation rate $\dot{\varphi}_\Mlabel{end} = 0.4 \,\omega_x$ and the interaction strength $\bar{g}N = 100$.
The trap is switched off at $\bar{T}_\Mlabel{off} = 16$ within the first period of the Bures distance $B(\tau)$ 
after the characteristic time $\bar{T}_\Mlabel{end} = 15$. 
The oscillation of the Bures distance stops after the trap is swiched off and it approaches its final value $B_\Mlabel{free}$. 
The black circles indicate the times for which the corresponding density distributions are displayed 
in Fig. \ref{fig:density-free}.}
\label{fig:free}
\end{center}
\end{figure}

The time dependence of the Bures distance for this case is depicted in Fig. \ref{fig:free}.
During the slow initial ramp, shown in Fig. \ref{fig:omega}, 
the Bures distance displays the collective oscillations discussed in subsection \ref{sec:num:rotation}. 
After switching off the trap at $\tau=T_\Mlabel{off}$, 
these oscillations stop and the Bures distance approaches its final value $B_\Mlabel{free}$. 
This behavior is a consequence of the fact that for a freely expanding BEC 
the adaptive matrix grows linearly for large times, as discussed in section \ref{sec:free-expansion-isotropic}, 
and thus the right-hand side of the affinely transformed GP equation~\eqref{ScaledGP} vanishes asymptotically.

\begin{figure}
 \begin{center}
  \includegraphics[scale=1]{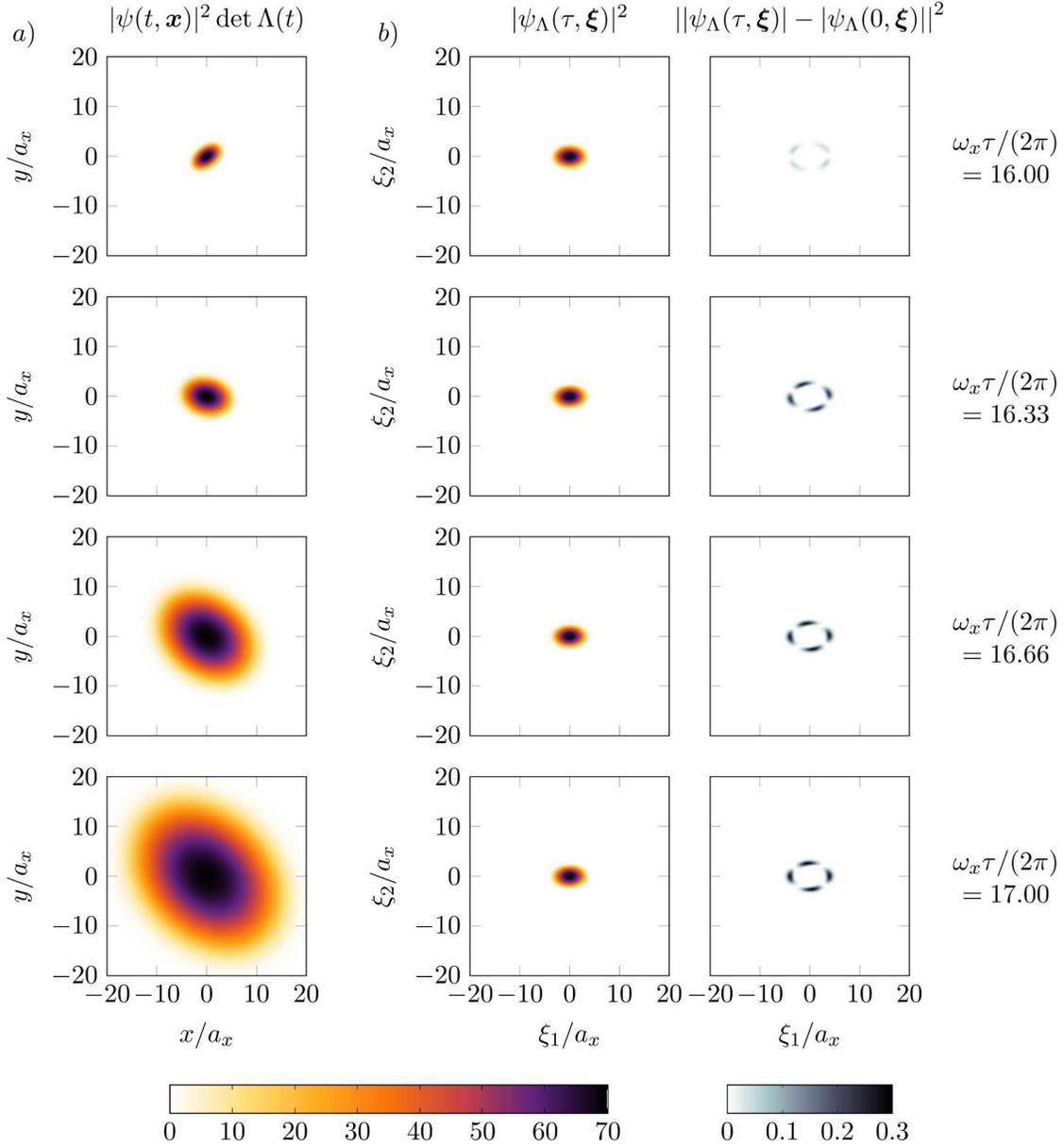}
  \caption{Density distributions $\abs{\psi(t,\rr)}^2$ and $\abs{\psi_\Mlabel{\Lambda}(\tau,\bzeta)}^2$ 
  together with the difference $\abs{\abs{\psi_\Mlabel{\Lambda}(\tau,\bzeta)}-\abs{\psi_\Mlabel{\Lambda}(0,\bzeta)}}^2$ 
  of a free expanding BEC after release from a rotating trap plotted for different times as indicated in Fig. \ref{fig:free}. 
  In original coordinates, Fig. \ref{fig:density-free} a), the rotation of the condensate comes to an end after an angle of about $\pi/2$, 
  while the size of the condensate grows continuously. In adapted coordinates, Fig. \ref{fig:density-free} b), no such effects are visible and the density distribution stays almost constant. 
  In the third column the difference $\abs{\abs{\psi_\Mlabel{\Lambda}(\tau,\bzeta)}-\abs{\psi_\Mlabel{\Lambda}(0,\bzeta)}}^2$ 
  between the time evolved state $\psi_\Mlabel{\Lambda}(\tau,\bzeta)$ and the initial state $\psi_\Mlabel{\Lambda}(0,\bzeta)$ 
  is illustrating the residual dynamics in adapted coordinates. 
  This difference only obtains non-zero values at the edge of the condensate 
  while its magnitude grows together with the Bures distance $B(\tau)$ shown in Fig. \ref{fig:free}. 
  Movie 2 displaying the whole time evolution is available in the online supplementary material.}
  \label{fig:density-free}
 \end{center}
 \end{figure}

The two-dimensional (non-integrated) density distributions
in the original coordinates $|\psi(t,\rr)|^2$, Fig. \ref{fig:density-free}a, as well as in the adapted coordinates 
$|\psi_\Mlabel{\Lambda}(\tau,\bzeta)|^2$, Fig. \ref{fig:density-free}b, 
are displayed at different times indicated by the black dots in Fig. \ref{fig:free}. The entire time evolution of these density distributions is presented in movie 2, which is available in the online supplementary material. 
Whereas the size of the BEC in the original coordinates quickly grows after the trap is switched off, 
the rotation of the BEC stops completely after a further rotation of about $\pi/2$, 
in accordance with the irrotationality condition, as well as the energy and momentum conservation.
It is worth mentioning that our results are in good agreement with 
the theoretical consideration \cite{Edwards2002} and the experimental data \cite{Hechenblaikner2002}.

However, in adapted coordinates, Fig. \ref{fig:density-free}b, 
neither an increase of the size of the BEC 
nor a significant rotation of the density distribution can be observed. 
Only by looking at the residual dynamics visualized again 
by the difference $||\psi_\Mlabel{\Lambda}(\tau,\bzeta)|-|\psi_\Mlabel{\Lambda}(0,\bzeta)||^2$, 
minor changes of $\psi_\Mlabel{\Lambda}(\tau,\bzeta)$ 
become visible at the edge of the condensate. 
Thus, the time-dependent TF approximation used for the affine approach is an excellent tool to predict 
the free expansion of an initially rotating BEC as well.

\subsubsection{Dependence of the Bures distance on the setup parameters}
\label{sec:num:dependence-sim-parameters}

\begin{figure}[h]
\begin{center}
\includegraphics[scale=1]{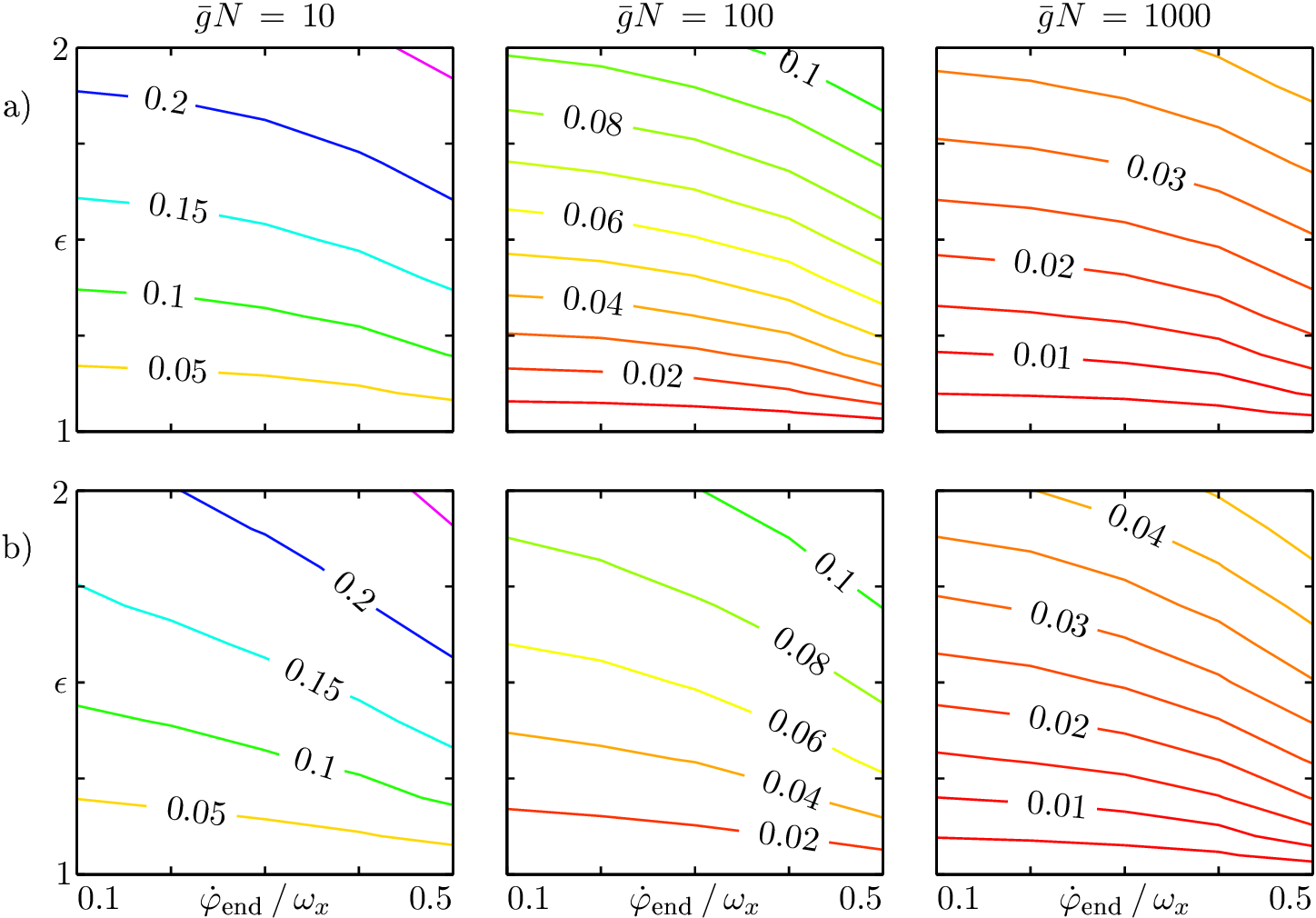}
\caption{Contour plot of (a) the maximal Bures distance $B_\Mlabel{rot}$ 
for a purely rotating trap with a time-dependent rotation rate as shown in Fig. \ref{fig:omega} 
and (b) the final Bures distance $B_\Mlabel{free}$ for a subsequent free expansion of an initially rotating trap 
as discussed in section \ref{sec:num:free-expansion}. 
For both scenarios the interaction strength $\bar{g}N$ increases from left to right, 
while the ranges of the anisotropy factor $\epsilon$ and 
the final rotation rate $\dot{\varphi}_\Mlabel{end}$ are the same for all plots, respectively. 
The differences between the case of a purely rotating trap (a) and a free expansion after the rotation (b) are minor. 
In general, an increase in the interaction strength $\bar{g}N$ leads to smaller values of the Bures distance, 
while in contrast the Bures distance takes on larger values as the anisotropy factor $\epsilon$ and 
the final rotation rate $\dot{\varphi}_\Mlabel{end}$ increase.}
\label{fig:sim}
\end{center}
\end{figure}

In extension of the two previous scenarios, we now provide a broader study of  
the accuracy of the time-dependent TF approximation.
In particular, we have solved numerically Eq. \eqref{ScaledGP} for a rotating BEC with and without 
subsequent free expansion for many different combinations 
of the interaction strength $\bar{g}N$, the anisotropy factor $\epsilon$ and the final rotation rate $\dot{\varphi}_\Mlabel{end}$,
thereby using $\bar{T}_\Mlabel{end} = 15$ as before. 
The results for the Bures distances $B_\Mlabel{rot}$ and $B_\Mlabel{free}$ are presented in Fig. \ref{fig:sim}.

As displayed in Fig. \ref{fig:sim}a the Bures distance $B_\Mlabel{rot}$ 
decreases with an increase of the interaction strength $\bar{g}N$ and 
increases for larger values of the anisotropy factor $\epsilon$ and 
the maximal rotation rate $\dot{\varphi}_\Mlabel{end}$. These observations can be explained as follows.

(i) With an increase of $\bar{g}N$, the interaction energy term in Eq. \eqref{ScaledGP} becomes more dominant compared to 
the kinetic energy term, resulting in an improved accuracy of the time-dependent TF approximation. 
For this reason, the Bures distance $B_\Mlabel{rot}$ decreases as the interaction strength $\bar{g}N$ grows.

(ii) Since a spherically symmetric trap with $\epsilon = 1$ cannot transfer any angular momentum to a BEC, 
the wave function $\psi_\Mlabel{\Lambda}(\tau,\bzeta)$ will remain in the initial state 
$\psi_\Mlabel{\Lambda}(0,\bzeta)$, giving rise to $B_\Mlabel{rot} = 0$. However, for $\epsilon > 1$ 
the rotation of the trap affects the dynamics of the BEC. 
The Bures distance $B_\Mlabel{rot}$ increases with an increase of $\epsilon$, 
since the kinetic energy term in Eq. \eqref{ScaledGP} depends on the spatial derivatives 
of the wave function and therefore increases for a stronger confinement along one axis. 
Thus, the time-dependent TF approximation works better for a system with small anisotropies. 

(iii) For larger values of the maximal rotation rate $\dot{\varphi}_\Mlabel{end}$, the total energy transfered from the trap to 
the BEC increases, raising the chance to excite collective modes of the condensate. 
However, these modes are not taken into account within the framework of the TF approximation. 
As a result, the amount of residual dynamics grows with the increase of the maximal rotation rate $\dot{\varphi}_\Mlabel{end}$. 
The quadrupole mode is well known to be excited when $\dot{\varphi}_\Mlabel{end}$ 
approaches the critical value $\omega_\Mlabel{c} = 0.71 \,\omega_x$ \cite{madison-2001,hodby-2001} and 
the time-dependent TF approximation breaks down completely in this regime. 
We have verified this fact with our numerical simulations and for the results presented here, 
we have always made sure to stay well below the threshold for vortex creation.

Since, the dependencies of $B_\Mlabel{rot}$ and $B_\Mlabel{free}$ on the simulation parameters are very similar to each other, 
as shown in Fig. \ref{fig:sim}, we conclude here that the subsequent free expansion 
of an initially rotating BEC does not substantially amplify the residual time evolution in the adapted coordinates. 

In summary, we find that the accuracy of the time-dependent TF approximation depends on various parameters 
with the most important one being the interaction strength $\bar{g}N$. For values $\bar{g}N \geq 100$, 
the transformed wave function $\psi_\Mlabel\Lambda(\tau,\bzeta)$ is practically constant in time and 
the time-dependent TF approximation holds true. 
In most experimental setups, one typically aims at a large number of particles in a BEC 
to improve the signal-to-noise ratio. 
For this reason, the TF approximation provides a valuable tool to describe the dynamics of a large class of rotating BECs.

\section{Constants of motion}\label{chap-constants-of-motion}

In the preceding section we have confirmed the accuracy of our method based on 
the time-dependent TF approximation by dedicated numerical simulations of
the affinely transformed GP equation \eqref{ScaledGP}. However, the reliability
of such numerical simulations themselves must also be guaranteed. One possible way
to test it is based on simulations of suitable scenarios that possess certain underlying
symmetries which by Noether's theorem give rise to specific constants of motion. 
In this chapter, we focus on the time-translational 
invariance and the rotational symmetry and discuss the corresponding energy and angular momentum 
conservation for both, the matrix differential equation~\eqref{MatrixDGL}
and the GP equation~\eqref{GPequation}.

In this section, we first show how the matrix differential equation \eqref{MatrixDGL} can be rewritten in terms of a Hamiltonian formalism. 
Second, we analyze the center-of-mass motion of the BEC with regard to a generalized version of the Ehrenfest 
theorem~\cite{Bodurov1998} and third, we present expressions for specific constants of motion of 
the GP equation~\eqref{GPequation} valid within the time-dependent TF approximation.

\subsection{Hamiltonian formalism for the matrix differential equation \label{sec:Hamilton:Formalism}}

Here we establish a connection between the matrix differential equation~\eqref{MatrixDGL} and 
the corresponding Hamiltonian formalism. For this purpose, we first perform a transformation of the adaptive matrix and the time coordinate 
to bring the matrix differential equation~\eqref{MatrixDGL} into a specific form that does no longer contain
the initial trap matrix $\Omega^2(0)$. We call the resulting equation the canonical form of the matrix differential equation. 
We then introduce the Hamiltonian and verify that the corresponding Hamilton equations of motion 
are equivalent to the canonical form of the matrix differential equation. 
With the help of this Hamiltonian we finally obtain two constants of motion for the matrix differential equation, 
which we later on relate to the conservation of the total energy and the angular momentum of the BEC.

\subsubsection{Canonical form of the matrix differential equation}

We start by defining the orthogonal matrix $O$ and the diagonal matrix $D$ via the diagonalization of the initial trap matrix $\Omega^2(0)$ according to
\begin{equation}
 O\,D\,O^\tp \equiv \Omega^2(0) \,. \label{matrix O}
\end{equation}
Moreover, we introduce a new time scale with the help of the quantity
\begin{equation}
  \alpha \equiv \left[\det\Omega^2(0)\right]^{\frac{1}{2d}} \,.  \\
\end{equation}
Based on these new quantities, we now consider the transformation
\begin{eqnarray}\label{eq:Trafo:Lambda}
 \Lambda & = & \frac{1}{\alpha}  \,O \, \tilde{\Lambda} \, D^{\frac{1}{2}} \, O^\tp\,, \nonumber \\
 t & = & \frac{1}{\alpha}\,\tilde{t}
\end{eqnarray}
to the new adaptive matrix $ \tilde{\Lambda}$ and the dimensionless time $\tilde{t}$.
The canonical form of the matrix differential equation is simply obtained by inserting the transformation~\eqref{eq:Trafo:Lambda}
into the matrix differential equation~\eqref{MatrixDGL} thereby taking into account the relation $\det(\Lambda)=\det(\tilde\Lambda)$. It describes the time evolution of the new adaptive matrix $\tilde{\Lambda}(\tilde t)$ and reads 
\begin{equation}
 \label{eq:Canonical:MatrixDGL}
 \tilde\Lambda^\tp(\bar t) \left[ \diffz{\tilde\Lambda}{\tilde t} + \tilde\Omega^2(\tilde t)\tilde\Lambda(\tilde t) \right]
 = \frac{\mathds{1}}{\det{\tilde\Lambda(\tilde t)}} \,,
\end{equation} 
where we have introduced the transformed trap matrix
\begin{equation}
 \tilde\Omega^2(\tilde t)\equiv \frac{1}{\alpha^2}\, O^\tp \Omega^2(\tilde t/\alpha)\, O.
\end{equation} 
The initial conditions~\eqref{InitialMatrix} accordingly transform to
\begin{equation}\label{eq:Canonical:InitialMatrix}
 \tilde\Lambda(0)  =  \alpha\, D^{-\frac{1}{2}} \quad\mbox{and}\quad 
 \diff{\tilde{\Lambda}}{\tilde t}\Big|_{\tilde t=0}  = 0 \,,
\end{equation}
whereas the irrotationality condition~\eqref{eq:IrrotCondition} preserves its form under the transformation~\eqref{eq:Trafo:Lambda} 
\begin{equation}
 \label{eq:Canonical:IrrotCondition}
  \tilde\Lambda^\tp(\tilde t) \diff{\tilde\Lambda}{\tilde t} = \diff{\tilde\Lambda^\tp}{\tilde t} \tilde\Lambda(\tilde t) \,.
\end{equation}
Next we show that the canonical form of the matrix differential equation~\eqref{eq:Canonical:MatrixDGL} 
can be embedded in a Hamiltonian formalism.

\subsubsection{Hamiltonian and equations of motion}

We start by introducing the momenta
\begin{equation}
 \label{eq:def:conjugate-variables-Pi}
  \tilde\Pi_{\alpha\beta} = \diff{\tilde\Lambda_{\alpha\beta}}{\tilde t}
\end{equation}
as conjugate variables to the matrix elements $\tilde\Lambda_{\alpha\beta}$ together with the Hamiltonian
\begin{equation}\label{eq:Hamilton:function}
 H(\tilde\Lambda,\tilde\Pi,\tilde t) = \frac{1}{2}\,
 \tr{\tilde\Pi^\tp\tilde\Pi + \tilde\Lambda^\tp \tilde\Omega^2(\tilde t)\tilde\Lambda} + \frac{1}{\det{\tilde\Lambda}}\,.
\end{equation}
The Hamilton equations of motion
\begin{eqnarray}
 \label{eq:Hamilton:equations-of-motion}
  \diff{\tilde\Lambda_{\alpha\beta}}{\tilde t} & = &  \{H, \tilde\Lambda_{\alpha\beta} \}_{(\tilde\Lambda,\tilde\Pi)}
  = \frac{\partial H}{\partial \tilde\Pi_{\alpha\beta}}= \tilde\Pi_{\alpha\beta} \,, \nonumber \\ 
  \diff{\tilde\Pi_{\alpha\beta}}{\tilde t} & = & \{H, \tilde\Pi_{\alpha\beta} \}_{(\tilde\Lambda,\tilde\Pi)} = - \frac{\partial H}{\partial \tilde\Lambda_{\alpha\beta}}
  = - \tilde\Omega^2\tilde\Lambda+\frac{1}{\det{\tilde\Lambda}}\,\tilde\Lambda^{-\tp}
\end{eqnarray}
are mathematically equivalent to the canonical form of the matrix differential equation \eqref{eq:Canonical:MatrixDGL}. 
Here, we have introduced the corresponding Poisson bracket 
\begin{equation}
 \label{eq:def:Poisson-brackets}
 \{f, g \}_{(\tilde\Lambda,\tilde\Pi)}\equiv 
 - \sum_{\alpha,\beta} \left( \pdiff{f}{\tilde\Lambda_{\alpha\beta}} \pdiff{g}{\tilde\Pi_{\alpha\beta}} - 
 \pdiff{g}{\tilde\Lambda_{\alpha\beta}} \pdiff{f}{\tilde\Pi_{\alpha\beta}} \right)
\end{equation}
of two functions $f$ and $g$ that depend on the matrix elements $\tilde\Lambda_{\alpha\beta}$ and $\tilde\Pi_{\alpha\beta}$
and made use of the relation
\begin{equation}
 \frac{{\rm d}}{{\rm d}\tilde\Lambda_{\alpha\beta}}\left(\det\tilde\Lambda\right) 
 = \det\tilde\Lambda \cdot \tr{\tilde\Lambda^{-1} \frac{\partial \tilde\Lambda}{\partial \tilde\Lambda_{\alpha\beta}}}  
 = \tilde\Lambda^{-\tp}_{\alpha\beta}\,\det\tilde\Lambda \,.
\end{equation}

When we accordingly define the ``momenta'' $\Pi\equiv \mathrm{d}\Lambda /\mathrm{d} t$, Eq.~\eqref{eq:Trafo:Lambda} implies the following 
transformation of the conjugate variables 
\begin{eqnarray}
 \label{eq:trafo:lambda-bar}
 \tilde\Lambda & = & \alpha\, O^\tp \, \Lambda\, O\, D^{-\frac{1}{2}} \,, \nonumber \\
 \tilde\Pi & = & O^\tp\, \Pi\, O\, D^{-\frac{1}{2}} \,.
\end{eqnarray}
We emphasize that Eq.~\eqref{eq:trafo:lambda-bar}
is not a canonical transformation and that a corresponding Hamiltonian formalism for the original matrices 
$\Lambda$ and $\Pi$ does not exist, as can be seen by inserting the transformation~\eqref{eq:trafo:lambda-bar} 
into the expression for the Poisson-bracket~\eqref{eq:def:Poisson-brackets}. As a result, we arrive at
\begin{eqnarray}
 \label{eq:results:Poisson-brackets}
 \{f, g \}_{(\tilde\Lambda, \tilde\Pi)} & =  
 - \frac{1}{\alpha} \sum_{\mu,\nu,\lambda} \left( \pdiff{f}{\Lambda_{\mu\nu}} \Omega^2_{\nu\lambda}(0) \pdiff{g}{\Pi_{\mu\lambda}} - 
 \pdiff{g}{\Lambda_{\mu\nu}} \Omega^2_{\nu\lambda}(0) \pdiff{f}{\Pi_{\mu\lambda}} \right) \,,
\end{eqnarray}
which clearly shows that the structure of the Poisson bracket~\eqref{eq:def:Poisson-brackets} is not preserved by the transformation~\eqref{eq:trafo:lambda-bar}.
Thus, there exist no corresponding Hamilton equations for the original matrix differential equation~\eqref{MatrixDGL}.

\subsubsection{Energy conservation}
\label{sec:energy:conservation}

When we consider a time-independent trapping potential with the constant trap matrix 
$\tilde\Omega^2(\tilde t) = \tilde\Omega_\ast^2$, the energy is a constant of motion given by  
the Hamiltonian $H(\tilde\Lambda, \tilde\Pi)$, Eq. \eqref{eq:Hamilton:function}. In this spirit, we call the quantity
\begin{equation}
 \label{eq:ELambda-bar}
 \tilde E(\tilde\Lambda,\tilde\Pi;\tilde t) \equiv 
 \frac{1}{2} \,\tr{\tilde\Pi^\tp\,\tilde\Pi + \tilde\Lambda^\tp\, \tilde\Omega^2(\tilde t)\, \tilde\Lambda} + 
 \frac{1}{\det{\tilde\Lambda}}
\end{equation}
the total energy associated with the matrix differential equation in canonical form also for a time-dependent trap matrix $\tilde\Omega^2(\tilde t)$. 

Using Eq.~\eqref{eq:def:conjugate-variables-Pi} together with the inverse of the transformation~\eqref{eq:Trafo:Lambda} in Eq.~\eqref{eq:ELambda-bar}, 
the total energy associated with the matrix differential equation $E_\Lambda(t)=\tilde E(\tilde \Lambda,\tilde \Pi; \tilde t)$ reads in terms of the original adaptive matrix
\begin{equation}
 \label{eq:ELambda}
 E_{\Lambda}(t) = \frac{1}{2}\, 
 \tr{\bigg(\diff{\Lambda^\tp}{t}\diff{\Lambda}{t} + \Lambda^\tp(t)\,\Omega^2(t)\,\Lambda(t)\bigg) \Omega^{-2}(0)} + 
 \frac{1}{\det{\Lambda(t)}} \,,
\end{equation}
where $\Omega^2(t)=\alpha^2 \, O\, \tilde\Omega^2(\tilde t)\, O^\tp$ and $\Omega^{-2}(0) = O\, D^{-1}\, O^\tp$. The fact that the total energy~\eqref{eq:ELambda}
is indeed a constant of motion if $\Omega^2(t) = \Omega_\ast^2$ can be easily verified by taking the time derive of Eq,~\eqref{eq:ELambda} on both sides and inserting 
the matrix differential equation~\eqref{MatrixDGL} on the right hand side of the resulting equation.

\subsubsection{Angular momentum conservation}

For an isotropic harmonic potential with corresponding trap matrix $\tilde\Omega^2(\tilde t) = \tilde\omega^2(\tilde t) \,\mathds{1}$,
the angular momentum of a BEC is preserved. In order to see how this conservation law is related to the matrix 
differential equation~\eqref{MatrixDGL}, we study the quantity
\begin{equation}
 \label{eq:def:angular:momentum:matrix}
 \tilde L (\tilde\Lambda, \tilde\Pi)\equiv \tilde\Lambda\, \tilde\Pi^\tp - \tilde\Pi\, \tilde\Lambda^\tp
\end{equation}
which we denote as angular momentum matrix associated with the matrix differential equation in canonical form. 
The angular momentum matrix~\eqref{eq:def:angular:momentum:matrix} can be found with the help of Noethers theorem and the irrotationality condition~\eqref{eq:Canonical:IrrotCondition} which reads in terms of both conjugate variables $\tilde\Lambda^\tp\, \tilde\Pi - \tilde\Pi^\tp\, \tilde\Lambda=0$.

We now prove that this angular momentum matrix is preserved for an isotropic trapping potential. For this purpose, we 
take the total derivative of $\tilde L$ with respect to the time $\tilde t$ and obtain
\begin{equation}\label{eq:total:derivative:barL}
 \diff{\tilde L}{\tilde t} = \frac{\partial \tilde L}{\partial \tilde t}+ \{H, \tilde L\}_{(\tilde\Lambda,\tilde\Pi)},
\end{equation}
where the Poisson bracket $\{H, \tilde L\}_{(\tilde\Lambda,\tilde\Pi)}$ is defined by Eq. \eqref{eq:def:Poisson-brackets}. Since the angular momentum matrix 
$\tilde L$ does not explicitly depend on the time $\tilde t$, 
the partial derivative $\partial\tilde L / \partial \tilde t$ vanishes. With the Hamiltonian~\eqref{eq:Hamilton:function}, 
the evaluation of the Poisson bracket finally yields
\begin{equation}
 \label{eq:Poisson:brackets:HL}
 \diff{\tilde L}{\tilde t} = \tilde\Lambda\, \tilde\Lambda^\tp\, \tilde\Omega^2(\tilde t) - 
 \tilde\Omega^2(\tilde t)\, \tilde\Lambda\, \tilde\Lambda^\tp \,.
\end{equation}
Thus, the angular momentum matrix $\tilde L$ is a constant of motion if the matrix $\tilde\Lambda\, \tilde\Lambda^\tp\, \tilde\Omega^2(\bar t)$ is symmetric, 
which is indeed the case for an isotropic trap with $\tilde\Omega^2(\tilde t) = \tilde\omega^2(\tilde t) \,\mathds{1}$. 

When we insert the transformation~\eqref{eq:trafo:lambda-bar} into the definition~\eqref{eq:def:angular:momentum:matrix} and use the transformation law $L_\Lambda(t)=O\,\tilde L(\tilde \Lambda,\tilde \Pi)\,O^\tp$, we obtain the angular momentum matrix in terms of the original adaptive matrix
\begin{equation}
 \label{eq:angular:momentum:matrix:Lambda}
 L_\Lambda(t) = \alpha \left[\Lambda(t)\, \Omega^{-2}(0)\, \diff{\Lambda^\tp}{t} - 
 \diff{\Lambda}{t}\Omega^{-2}(0)\Lambda^\tp(t) \right] \,.
\end{equation}
We emphasize that the angular momentum matrix is antisymmetric and explicitly depends on the initial trapping potential.

\subsection{Constants of motion of the Gross-Pitaevskii equation}

Our efficient description of the time evolution of a rotating BEC within the TF regime, Eq.~\eqref{eq:solutionTF} rests upon two different, but related approximations:
(i) the initial macroscopic wave function $ \psi\left(0,\rr\right)$ is assumed to be given by the ground state in the TF approximation, Eq.~\eqref{eq:initialTFwavefunc},
and (ii) the kinetic energy term in the affinely transformed GP equation~\eqref{ScaledGP} is supposed to be negligible in comparison to the potential and interaction energies involved. According to this second assumption known as the time-dependent TF approximation, the affinely transformed wave function, 
being the solution of Eq. \eqref{ScaledGP}, does not display any essential dynamics
\begin{equation}\label{eq:general:notimeevolution}
 \psi_\Mlabel\Lambda(\tau,\bzeta) \approx \psi_\Mlabel\Lambda(0,\bzeta).
\end{equation} 
In section~\ref{sec:effective_description} we used a combination of both approximations, (i) and (ii), to derive the efficient description of the BEC dynamics within the TF regime given by Eq.~\eqref{eq:solutionTF}. 

However, for BECs that are not deep within the TF regime and whose ground state differs significantly from Eq.~\eqref{eq:initialTFwavefunc}, the assumption~\eqref{eq:general:notimeevolution} can still provide very accurate results for some scenarios such as freely expanding BECs. 
For this reason, we first seek an efficient description that does not involve approximation (i), but includes the exact ground state $\psi(0,\rr)$ of a BEC or its numerically determined approximation. Based on this slightly more general description, we derive approximate expressions for the total energy and the angular momentum of a BEC that allow further tests of assumption~\eqref{eq:general:notimeevolution} by analyzing possible time dependencies of the constants of motion that should not occur due to the chosen symmetry of the GP equation. Eventually, we also include approximation (i) and relate the resulting expressions for the total energy and the angular momentum based on the TF ground state~\eqref{eq:initialTFwavefunc} to the corresponding constants of motion associated with the matrix differential equation.

\subsubsection{Effective description of the dynamics based solely on the time-dependent Thomas-Fermi approximation}

In order to determine the slightly generalized efficient description of the BEC dynamics, we first relate
the initial state in adapted coordinates $\psi_\Mlabel\Lambda(0,\bzeta)$ with the original ground state $\psi(0,\rr)$ 
by evaluating Eq.~\eqref{eq:volleInverseTrafo} at $t = 0$ giving rise to
\begin{equation}
 \label{eq:general:initial:state}
 \psi_\Mlabel\Lambda(0,\bzeta) = \Me^{-\frac{\Mi}{\hbar} \calP(0)\,\left[\bzeta + \frac{1}{2} \calR(0)\right]}\, 
 \psi\left(0, \bzeta + \calR(0)\right)\,.
\end{equation}
Next, we obtain $\psi_\Mlabel\Lambda(\tau,\bzeta)$ simply by combining Eqs.~\eqref{eq:general:notimeevolution} 
and~\eqref{eq:general:initial:state}. We insert the result into Eq.~\eqref{eq:volleTrafo} and thereby use the 
substitution $\bzeta = \Lambda^{-1}(t)\left[\rr - \calR(t)\right]$. Hence, we find as efficient description of the 
BEC dynamics that is solely based on the time-dependent TF approximation 
\begin{equation}
 \label{eq:general:EffectiveTimeEvolTF}
  \psi(t,\rr) \approx 
  \frac{1}{\sqrt{\det\Lambda(t)}}\;
  \Me^{\Mi\,\chi(t,\rr)}\;
  \psi\left(0, \Lambda^{-1}(t)\,\left[\rr - \calR(t)\right]+\calR(0)\right)\,,
\end{equation}
where we have introduced the phase 
\begin{eqnarray}
 \label{eq:general:PhaseFactor}
 \hspace{-2cm} \chi(t,\rr) =
   \frac{1}{\hbar}\, & \left\{ \calS_2(t) - \beta(t) +\calP(t)\,\rr +
   \frac{m}{2} \left[\rr-\calR(t)\right]^\tp\, C(t) \left[\rr-\calR(t)\right]\right. \; \nonumber \\
   & \left. - \calP(0)\left(\Lambda^{-1}(t)\left[\rr-\calR(t)\right] + \calR(0)\right)\right\}. 
\end{eqnarray}
Note that by insertion of $\psi(0,\rr)=\phi_\Mlabel{TF}(\rr)$ together with Eq.~\eqref{eq:initialTFwavefunc} into Eq.~\eqref{eq:general:EffectiveTimeEvolTF},
we consequentially arrive at the expression~\eqref{eq:solutionTF}.

\subsubsection{Generalized Ehrenfest Theorem for a Bose-Einstein condensate in a harmonic trap}
\label{sec:Ehrenfest}

According to the generalized Ehrenfest-Theorem \cite{Bodurov1998}, the expectation values
\begin{eqnarray}
 \ave{\hat\rr}_{\psi(t)} 
 &\equiv \frac{1}{N} \int\limits_{\mathds{R}^d}\psi^*(t,\rr)\,\rr\,\psi(t,\rr)\,\Mdiff^d x \,,
\label{aveposition}\\
 \ave{\bpp}_{\psi(t)} &\equiv 
 \frac{1}{N} \int\limits_{\mathds{R}^d}\psi^*(t,\rr)\,\bpp\,\psi(t,\rr)\,\Mdiff^d x 
\label{avemomentum}
\end{eqnarray}
of the position $\hat\rr = \rr$ and the momentum $\bpp\equiv -\Mi\hbar\bnabla_{\rr}$ operators satisfy the classical equations of motion
\begin{eqnarray}
 \label{eq:averages_equationsOfMotion}
 \diff{}{t}\ave{\hat\rr}_{\psi(t)} & = & \frac{1}{m}\ave{\bpp}_{\psi(t)} \,, \nonumber\\
 \diff{}{t}\ave{\bpp}_{\psi(t)} & = & -m\,\Omega^{2}(t)\left[\ave{\hat\rr}_{\psi(t)} - \brho(t)\right]+ \bgg(t)\,.
\end{eqnarray}
The expectation values are evaluated with the full solution $\psi(t,\rr)$ of the GP equation~\eref{GPequation} for the harmonic potential, Eq. \eqref{eq:potential}.
Since the Eqs. \eqref{eq:averages_equationsOfMotion} coincide with the Eqs.~\eref{eq:equationsOfMotion} 
for the time-dependent vectors $\calR(t)$ and $\calP(t)$, we can identify the latter ones 
with the center-of-mass position $\ave{\hat\rr}_{\psi(t)}$ and the average momentum $\ave{\bpp}_{\psi(t)}$
by simply choosing the initial conditions according to
\begin{equation}
 \label{eq:RP_initial_conditions}
  \calR(0)=\ave{\hat\rr}_{\psi_0}
  \qquad \mbox{and} \qquad
  \calP(0)=\ave{\bpp}_{\psi_0}\,.
\end{equation}
Here, we have introduced the shorthand notation of the exact initial state $\psi_0\equiv \psi(0,\rr)$. 
We emphasize that the parameters $\calR(t)$ and $\calP(t)$ introduced in section~\ref{sec:com_Elimination},
have strictly speaking no physical meaning unless their initial conditions are fixed in accordance 
with Eq.~\eqref{eq:RP_initial_conditions}. 

Next, we discuss if the above assignment 
\begin{equation}
 \label{eq:RP_equal_expectationValues}
  \calR(t)=\ave{\hat\rr}_{\psi(t)}
  \qquad \mbox{and} \qquad
  \calP(t)=\ave{\bpp}_{\psi(t)}
\end{equation}
is also consistent with our efficient description, Eq.~\eqref{eq:general:EffectiveTimeEvolTF}, based on the time-dependent TF approximation. 
For this purpose, we insert Eq.~\eqref{eq:general:EffectiveTimeEvolTF} into the Eqs. \eqref{aveposition} and \eqref{avemomentum} 
and make use of the substitution 
\begin{eqnarray} 
 \label{eq:Ehrenfest_substitutions}
  \rr' & = & \Lambda^{-1}(t)\left[\rr-\calR(t)\right]+\calR(0)
\end{eqnarray}
to arrive at the expectation values 
\begin{eqnarray}
  \hspace{-1cm}\ave{\hat\rr}_{\psi(t)}
  &\approx \calR(t) + \Lambda(t)\big[\ave{\hat\rr}_{\psi_0} - \calR(0)\big], \label{eq:position_approx:expect} \\
  \hspace{-1cm}\ave{\bpp}_{\psi(t)}
  &\approx\calP(t)
  + \Lambda^{-\tp\!}(t)\big[
    \ave{\bpp}_{\psi_0} - \calP(0)
  \big]
  + m\,\diff{\Lambda}{t}\big[
  	\ave{\hat\rr}_{\psi_0} - \calR(0)
  \big] \,. \label{eq:momentum_approx:expect}
\end{eqnarray} 
Thus, using $\calR(0)$ and $\calP(0)$ according to Eq.~\eqref{eq:RP_initial_conditions}, 
the last two expressions simply reduce to Eqs.~\eqref{eq:RP_equal_expectationValues}, 
which proves that our efficient description based on the time-dependent TF approximation, Eq.~\eqref{eq:general:EffectiveTimeEvolTF}, 
satisfies the generalized Ehrenfest theorem.

\subsubsection{Energy of a Bose-Einstein condensate}
\label{sec:energy:TF:approximation}

Here we use the approximate BEC wave function $\psi(t,\rr)$ given by Eq. \eqref{eq:general:EffectiveTimeEvolTF} to derive an explicit expression for the total energy 
\begin{equation}\label{eq:totEnergy}
E(t) = E_\Mlabel{kin}(t) + E_\Mlabel{pot}(t) + E_\Mlabel{int}(t)
\end{equation}
of the BEC within the time-dependent TF approximation, where 
the kinetic, the potential and the interaction energy are defined by
\begin{eqnarray}
 E_\Mlabel{kin} (t)
& = & \frac{\hbar^2}{2m} \int_{\mathds{R}^d}
  |{\bnabla_\mboxbf{x}} \psi(t,\mboxbf{x})|^2\,\Mdiff^d x = 
  \frac{N}{2m} \ave{\bpp^2}_{\psi(t)}, \label{eq:kinEnergy:def} \\
E_\Mlabel{pot} (t) & = & \int_{\mathds{R}^d} V(t,\rr) \,|\psi(t,\mboxbf{x})|^2\, \Mdiff^d x 
 = N \ave{V(t,\rr)}_{\psi(t)}, \label{eq:potEnergy:def} \\
E_\Mlabel{int} (t) & = & \frac{g}{2} \int_{\mathds{R}^d}|\psi(t,\mboxbf{x})|^4\, \Mdiff^d x
  = \frac{Ng}{2} \ave{|\psi(t,\mboxbf{x})|^2}_{\psi(t)}.\label{eq:intEnergy:def}
\end{eqnarray}

In \ref{sec:appEnergy} we evaluate each of these energy terms and obtain as the energy per particle
\begin{eqnarray}
 \label{eq:totalEnergy:all:terms}
 \fl \frac{E(t)}{N}
  = \frac{\calP^2(t)}{2m}
  + V(t,\calR(t)) \nonumber 
  + \frac{1}{2m} \,\mathrm{Tr} \left[\Lambda^{-1}(t)\,\Lambda^{-\tp}(t) \,\ave{\bpp \otimes \bpp^\tp}_{|\psi_0|} \right]
  + \frac{E_\Mlabel{int}(0)}{N \det\Lambda(t)}  \nonumber \\
 \fl \qquad\quad
  + \frac{m}{2}\, \mathrm{Tr} \left\{\left[\diff{\Lambda^\tp}{t}\diff{\Lambda}{t} + \Lambda^\tp(t)\Omega^2(t)\Lambda(t) \right] 
 \ave{\left[\hat\rr - \calR(0)\right] \otimes \left[\hat\rr - \calR(0) \right]^\tp }_{\psi_0} \right\}  \,,  
\end{eqnarray}
where we have assumed that the initial wave function $\psi(0,\rr)$ is given by
\begin{equation}\label{eq:initial:wave:function:inertial:frame}
 \psi(0,\rr) = \phi_\Mlabel{D}\left(\rr - \calR(0)\right) \, \Me^{\frac{\Mi}{\hbar}\calP(0)\,\rr} \,.
\end{equation}
Here, $\phi_\Mlabel{D}(\rr)$ is the real-valued ground state of the BEC in the comoving frame of the trap as discussed in section \ref{sec:rho} and the phase $\Mi\calP(0)\,\rr/\hbar$ accounts for the initial momentum $\calP(0)$ of the BEC in the inertial frame of reference $\rr$, analogously to the TF ground state \eqref{eq:initialTFwavefunc}.
The first two terms in Eq. \eqref{eq:totalEnergy:all:terms} display the kinetic and the potential energy 
of the center of mass of the BEC, respectively, whereby we have used $V(t,\calR(t))$ as shorthand notation for the harmonic potential~\eqref{eq:potential}.
The third term corresponds to the quantum pressure and its expectation value is evaluated with respect to the absolute value $|\psi(0,\rr)|$ of the initial wave function. 
The forth term describes the mean-field interaction energy. 
The last term is determined by the explicit form of the initial wave function $\psi(0,\rr)$ and 
represent the inner dynamics of the condensate associated with the deformation of the BEC which corresponds to the evolution of the adaptive matrix $\Lambda(t)$.

When the initial wave function $\psi(0,\rr)$ is given by the TF ground state 
$\phi_\Mlabel{TF}(\rr)$, Eq. \eqref{eq:initialTFwavefunc}, the energy per particle reads
\begin{equation}
 \label{eq:totalEnergy_TF_approx}
  \frac{E_\Mlabel{TF}(t)}{N} = \frac{\calP^2(t)}{2m}  + V(t,\calR(t)) + \frac{2}{d+4}\,\mu_\Mlabel{TF}\, E_\Lambda (t)
\end{equation}
as shown in \ref{sec:appEnergy:TF:groundState}. Here $E_\Lambda (t)$ denotes the total energy associated with the matrix differential equation given by Eq. \eqref{eq:ELambda} whereas $\mu_\Mlabel{TF}$ is the chemical potential within the TF approximation, Eq. \eqref{eq:chemicalPotentialTF}. 

By applying  Eq. \eqref{InitialMatrix} to the energy per particle \eqref{eq:totalEnergy_TF_approx} and assuming without loss of generality that $V(0,\calR(0)) = 0$, 
the initial energy per particle for a BEC in the TF ground state $\phi_\Mlabel{TF}(\rr)$ reads
\begin{equation}
 \label{eq:totalEnergy_TF_approx_initial}
  \frac{E_\Mlabel{TF}(0)}{N} = \frac{\calP^2(0)}{2m}  +  \frac{d+2}{d+4}\,\mu_\Mlabel{TF}
\end{equation}
and is determined solely by the initial momentum $\calP(0)$ and the chemical potential $\mu_\Mlabel{TF}$.

\subsubsection{Angular momentum of a Bose-Einstein condensate}
\label{sec:angular:momentum:TF}

In analogy to the previous subsection, we now discuss the angular momentum of a BEC within the time-dependent TF approximation. 
Indeed, as derived in \ref{sec:appAngularMomentumOperator:original:coordinates} the angular momentum operator with respect to the center-of-mass coordinates $\calR(t)$ and $\calP(t)$ is defined by
\begin{equation}
 \label{eq:def:angular:momentum:center_of_mass}
  \hat\Mbf{L} = \big(\hat\rr-\calR(t)\big) \times \big(\bpp-\calP(t)\big) \,.
\end{equation}

In \ref{sec:appAngularMomentum:expect:value} we show that the expectation value of the angular momentum operator, Eq. \eqref{eq:def:angular:momentum:center_of_mass}, evaluated with respect to the approximate wave function \eqref{eq:general:EffectiveTimeEvolTF}, is given by the relation
\begin{equation}
 \label{eq:angular:momentum:expectation:value}
  \ave{\hat\Mbf{L}}_{\psi(t)} = \ave{\left\lbrace \Lambda(t)\left[\hat\rr - \calR(0)\right] \right\rbrace 
  \times \left\lbrace m \,\diff{\Lambda}{t}\, \left[\hat\rr - \calR(0)\right] + \Lambda^{-\tp} \,\bpp\right\rbrace}_{\psi_0} \,,
\end{equation}
where we have used the initial conditions \eqref{eq:RP_initial_conditions}. Here, the first vector of the cross product represents the position, while the second vector corresponds to the momentum associated with the inner dynamics of the BEC. 
For the sake of a concise mathematical notation, we map the angular momentum operator to the so-called angular momentum matrix operator via $\hat{L}_{kl} = \epsilon_{jkl} \hat{L}_j$, where $\varepsilon_{jkl}$ denotes the Levi-Civita symbol. Applied to Eq. \eqref{eq:angular:momentum:expectation:value} this mapping yields the following expression for the matrix operator
\begin{eqnarray}
 \label{eq:angular:momentum:tensor:expect:value}
 \fl\ave{\hat{L}}_{\psi(t)} = m \left[\Lambda(t)\ave{\left[\hat\rr - \calR(0)\right] \otimes 
 \left[\hat\rr - \calR(0)\right]^\tp}_{\psi_0} \diff{\Lambda^\tp}{t} \right. \nonumber \\
 \fl\qquad\qquad\quad \left. - \diff{\Lambda}{t}\ave{\left[\hat\rr - \calR(0)\right] \otimes 
 \left[\hat\rr - \calR(0)\right]^\tp}_{\psi_0} \Lambda^\tp(t)\right] \nonumber \\
 \fl\qquad\qquad + \Lambda(t) \ave{\left[\hat\rr - \calR(0)\right] \otimes \bpp^\tp}_{\psi_0} \Lambda^{-1}(t) - \Lambda^{-\tp}(t) \ave{\bpp \otimes \left[\hat\rr - \calR(0)\right]^\tp}_{\psi_0} \Lambda^\tp(t) \,.
\end{eqnarray}
This matrix notation of the angular momentum is especially useful if the initial wave function $\psi(0,\rr)$ is approximated by the TF ground state $\phi_\Mlabel{TF}(\rr)$, Eq. \eqref{eq:initialTFwavefunc}, since in that case Eq. \eqref{eq:angular:momentum:tensor:expect:value} reduces to
\begin{equation}
 \label{eq:angular:momentum:tensor:TF:ground:state}
  \ave{\hat{L}}_\Mlabel{TF} = \frac{2}{d+4} \frac{\mu_\Mlabel{TF}}{\left[\det\Omega^2(0)\right]^{\frac{1}{2d}}} \; L_\Lambda (t)
\end{equation}
as discussed in \ref{sec:appAngularMomentum:TF:ground-state} with the angular momentum matrix $L_\Lambda(t)$ given by Eq.~\eqref{eq:angular:momentum:matrix:Lambda}. 

In conclusion, the deep connection between the constants of motion of the GP equation, Eq. \eqref{GPequation}, and the matrix differential equation, Eq. \eqref{MatrixDGL}, illustrated by Eqs. \eqref{eq:totalEnergy_TF_approx} and \eqref{eq:angular:momentum:tensor:TF:ground:state}, once more highlights the benefit of the affine approach for the description of the BEC dynamics. Not only grants the solution of Eq. \eqref{MatrixDGL} valuable insights into the time evolution of a BEC, but it is also relevant for numerical applications.

\section{Special properties of free expanding Bose-Einstein condensates}
\label{sec:free-expansion-properties}
Experiments dealing with quantum gases usually involve time-of-flight pictures to deduce the state of the system under study. It is well known, and we will briefly recall this statement later, that in the long-time limit the spatial density distribution obtained by time-of-flight measurements is determined by the initial momentum distribution of the quantum gas if the free time evolution is described by the Schr\"odinger equation. In other words, a non-interacting quantum gas undergoes a ballistic expansion if it is released from the trapping potential. However, it is not obvious that this statement holds true for free expanding BECs governed by the non-linear GP equation. In fact, for BECs a reverse relation applies \cite{PhD_ME}, namely that the momentum distribution in the long-time limit is given by the initial spatial density distribution. This relation is not only relevant for the interpretation of time-of-flight pictures of BECs, but it also plays a role for matter-wave interferometry \cite{Kleinert2015}, where the visibility of the phase oscillations quickly decreases when the momentum distribution of the atomic clouds exceed the corresponding line width of the laser pulses.

This section starts with the discussion of the analytic solutions of the matrix differential equation \eqref{MatrixDGL} for freely expanding BECs in the long-time limit that start from isotropic initial traps. We then recall the long-time behavior for the Schr\"odinger equation before we turn to the GP equation.

\subsection{Analytic solutions of the matrix differential equation}
\label{sec:free-expansion-isotropic}
For a pure free expansion, that is $\Omega^2(t) = 0$ for all times $t>0$, the matrix differential equation \eqref{MatrixDGL} possesses analytical solutions that can be used to determine the macroscopic wave function of a BEC in the long-time limit. In order to derive these solutions we assume without loss of generality that the coordinate system $\rr$ is aligned with the principal axes of the trap at the initial time $t=0$. As a consequence, the adaptive matrix $\Lambda(t)$ is diagonal during the free expansion, that is $\Lambda(t)=\op{diag}\left[\lambda_1(t),\ldots,\lambda_d(t)\right]$ for $t>0$, and the scaling parameters $\lambda_i$ with $i=1,2,...,d$ are determined by the system of non-linear differential equations
\begin{equation}\label{eq:lambda_dgl_scalar}
 \diffz{\lambda_i}{t} = \frac{\omega^2_i(0)}{\lambda_i\,\prod_{j=1}^d \lambda_j}  \,,
\end{equation}
which are a special case of equations \eqref{MatrixDGL} or \eqref{eq:Castin-Kagan} for free evolution. Here the frequencies $\omega_i(0)$ denote the initial trap frequencies along the principal axes. The initial conditions \eqref{InitialMatrix} reduce to 
\begin{equation}\label{eq:lambda_initial_scalar}
\lambda_i(0) = 1 \quad\mbox{and}\quad
  \left.\diff{\lambda_i}{t}\right|_{t=0} = 0 \,, 
\end{equation}
accordingly.

\subsubsection{Long-time limit}
Since the scaling parameters $\lambda_i(t)$ determine the size of the condensate, 
they grow as the condensate expands freely. Hence, the right hand side of Eq. \eqref{eq:lambda_dgl_scalar} 
vanishes in the long-time limit and the scaling parameters approach the linear dependence
\begin{equation}\label{eq:long-time-lambda}
  \lambda_i(t) \cong a_i + b_i \,t 
\end{equation}
as $t \rightarrow \infty$. In general, the constants $a_i$ and $b_i$ are determined numerically. However, in the case of an isotropic initial trap Eq. \eqref{eq:lambda_dgl_scalar} can be solved analytically also yielding the values for $a_i$ and $b_i$ as shown in the next subsection.

\subsubsection{Isotropic case}
For an isotropic initial trap, that is $\omega_i(0) = \omega_0$, the scaling parameters $\lambda_i(t)$ are given by the single scaling parameter $\lambda(t)$ for all $i=1,2,...,d$ 
and the system of non-linear differential equations, Eq. \eqref{eq:lambda_dgl_scalar}, decouples. We show in \ref{app:Lambda:DGL:isotropic} that the function $\lambda(t)$ is a solution of the integral equation 
\begin{equation}
 \label{eq:integral-equation-lambda}
 \int_1^{\lambda(t)}d\zeta\,\frac{\zeta^{d/2}}{\sqrt{\zeta^d-1}}=\sqrt{\frac{2}{d}}\;\omega_0 t.
\end{equation}
Depending on the dimensionality $d$, Eq. \eqref{eq:integral-equation-lambda} gives rise to explicit, $d=2$, or implicit, $d=1$ and $d=3$, solutions for the scaling parameter $\lambda(t)$, which are discussed in the next paragraphs. 

\paragraph*{\bf The case $d=1$}
The left-hand side of Eq. \eqref{eq:integral-equation-lambda} gives rise to the transcendental equation
\begin{equation*}
\sqrt{\lambda (\lambda - 1)} + \ln\left(\sqrt{\lambda}+\sqrt{\lambda-1}\right)=\sqrt{2} \,\omega_0 t
\end{equation*}
which can also be written as
\begin{equation}
 \label{eq:solution:lambda:1D}
 \lambda=\frac{\sqrt{2} \,\omega_0 t}{\sqrt{1-\frac{1}{\lambda}}+\frac{1}{\lambda}\ln\left(\sqrt{\lambda}+\sqrt{\lambda-1}\right)}.
\end{equation}
It is clear that for long times, that is $\omega_0 t\gg 1$, the function $\lambda \cong \sqrt{2}\,\omega_0 t$, 
since the denominator in the right-hand side of Eq. \eqref{eq:solution:lambda:1D} approaches unity as $\lambda \rightarrow \infty$. 
Taking into account the terms in the denominator scaling as $\lambda^{-1}$ and $\ln(\lambda)\lambda^{-1}$, we finally arrive at the approximate solution
\begin{equation}
 \label{eq:long-time-1D}
 \lambda_{\mathrm{1D}}(t)\cong \sqrt{2} \,\omega_0 t-\frac{1}{2}\left[\ln\left(4\sqrt{2}\,\omega_0 t\right)-1\right]
\end{equation}
for $\omega_0 t\gg 1$. Here we have neglected all terms, that vanish in the long-time limit. 

\paragraph*{\bf The case $d=2$}
In the case of two dimensions, the integral in the left-hand side of Eq. \eqref{eq:integral-equation-lambda} 
yields $\sqrt{\lambda^2(t)-1}$, leading to the exact solution
\begin{equation}\label{eq:solution:lambda:2D}\label{eq:long-time-2D}
 \lambda_{\mathrm{2D}}(t) = \sqrt{1 + \omega_0^2\,t^2}
\end{equation}
valid for all $t \geq 0$.

\paragraph*{\bf The case $d=3$}
For a three dimensional BEC the integral in the left-hand side of Eq. \eqref{eq:integral-equation-lambda} 
can be expressed in terms of the Gaussian hypergeometric function $_2F_1(a,b,c;z)$ leading to the implicit equation 
\begin{equation}
 \label{eq:solution:lambda:3D}
 _2F_1\left(-\frac{1}{3},\frac{1}{2},\frac{2}{3};\frac{1}{\lambda^3}\right)\lambda-
 \frac{\sqrt{\pi}\Gamma\left(\frac{2}{3}\right)}{\Gamma\left(\frac{1}{6}\right)}=\sqrt{\frac{2}{3}} \,\omega_0 t
\end{equation}
for $\lambda(t)$, where $\Gamma(x)$ denotes the Gamma function. Since the hypergeometric function $_2F_1(-1/3,1/2,2/3;1/\lambda^3)$ approaches unity as $\lambda\rightarrow \infty$, 
the solution $\lambda(t)$ of Eq. \eqref{eq:solution:lambda:3D} in the long time limit, that is $\omega_0 t\gg 1$, is given by
\begin{equation}
 \label{eq:long-time-3D}
 \lambda_{\mathrm{3D}}(t) \cong \frac{\sqrt{\pi}\Gamma\left(\frac{2}{3}\right)}{\Gamma\left(\frac{1}{6}\right)}
 +\sqrt{\frac{2}{3}} \,\omega_0 t.
\end{equation}

When we compare the relations \eqref{eq:long-time-1D}, \eqref{eq:long-time-2D}, and \eqref{eq:long-time-3D} with 
Eq. \eqref{eq:long-time-lambda}, we find that the coefficient $b_i = \sqrt{2/d}\,\omega_0$.

\subsection{Connection between the long-time behavior of the wave function and the initial state}
Here we show how the long-time behavior of the macroscopic wave function of a BEC can be estimated 
in the case of free expansion, although the dynamics is governed by the non-linear GP equation. Since the approximate analytic result for BECs contrasts the long-time solution of non-interacting wave functions, we first have a look at the Schr\"odinger equation before discussing the GP equation. 

\subsubsection{Schr\"odinger equation}
The Schr\"odinger equation describing the free evolution of the non-interacting wave function $\psi(t,\rr)$ in a $d$-dimensional space reads
\begin{equation}
 \label{eq:schrodinger-equation-free}
 \Mi \hbar \,\frac{\partial \psi(t,\rr)}{\partial t}  = \frac{\bpp^2}{2m}\,\psi(t,\rr)
\end{equation}
and has the formal solution
\begin{equation}
 \label{eq:formal-solution-schroedinger}
 \psi(t,\rr) = \Me^{-\frac{\Mi}{\hbar} \frac{\bpp^2}{2m} \,t} \psi(0,\rr)
\end{equation}
in terms of the initial wave function $\psi(0,\rr)$. When we use the Fourier transformation
\begin{equation}\label{eq:Fourier-trafo-p-x}
 \tilde\psi(0,\Mbf{p}) \equiv \frac{1}{(2\pi\hbar)^{d/2}}\int_{\mathds{R}^d}\,\Me^{-\frac{\Mi}{\hbar}\Mbf{p} \rr}\,
 \psi(0,\rr)\,\Mdiff^d x
\end{equation}
of the wave function $\psi(0,\rr)$, we can cast Eq. \eqref{eq:formal-solution-schroedinger} into the form
\begin{equation}\label{eq:Schroedinger-step2}
 \psi(t,\rr) = \frac{1}{(2\pi\hbar)^{d/2}} \,\Me^{-\frac{\Mi}{\hbar} \frac{m}{2t} \rr^2} \int_{\mathds{R}^d} 
 \Me^{-\frac{\Mi}{\hbar} \frac{t}{2m} \left(\Mbf{p} - \frac{m\rr}{t}\right)^2} \,\tilde\psi(0,\Mbf{p})\,\Mdiff^d p \,.
\end{equation}

For long times, that is in the limit $t\rightarrow \infty$, the exponential function inside the integral on the right-hand side of Eq. \ref{eq:Schroedinger-step2} oscillates very rapidly and only the values of $\Mbf{p}$ in a narrow region around $\Mbf{p} = m\,\rr/t$ contribute to the integral. Hence, when we assume that the momentum distribution $\tilde\psi(0,\Mbf{p})$ is a smooth function of $\Mbf{p}$ in this region, we can evaluate $\tilde\psi(0,\Mbf{p})$ at $\Mbf{p} = m\,\rr/t$ and perform the remaining integral, leading to the relation
\begin{equation}
 \psi(t,\rr)\cong \left(\frac{\Mi m}{t}\right)^{d/2} \,\Me^{-\frac{\Mi}{\hbar} \frac{m}{2t} \rr^2} \,
 \tilde\psi\left(0,\frac{m\rr}{t}\right).
\end{equation}
Consequently, as is well known, the spatial density distribution in the long time limit
\begin{equation}
 \label{eq:Schrodinger-result}
 \left|\psi(t,\rr)\right|^2 \cong \left(\frac{m}{t}\right)^d \,\left|\tilde\psi\left(0,\frac{m\rr}{t}\right)\right|^2
\end{equation}
is determined by the shape of the initial momentum distribution $\tilde\psi(0,\Mbf{p})$.

\subsubsection{Gross-Pitaevskii equation}
For free expanding BECs a similar relation can be established based on the affine approach. In order to compare the results derived for the Schr\"odinger equation and the GP equation, respectively, it is suitable to consider the wave function $\psi_\Mlabel{D}(t,\rr)$ that describes the inner dynamics of the BEC where the center-of-mass motion has already been eliminated. As shown in section \ref{aff_tran}, the time evolution of $\psi_\Mlabel{D}(t,\rr)$ is governed by the transformed GP equation \eqref{GPequationD} which is related to the affinely transformed wave function $\psi_\Mlabel\Lambda(\tau,\bzeta)$ by Eq. \eqref{DefpsiLambda}. If the initial wave function $\psi_\Mlabel{D}(0,\rr)$ coincides with the TF ground state, Eq. \eqref{eq:initialTFwavefuncLambda}, the time-dependent TF approximation \eqref{eq:Notimeevol_for_trans_wavefct} holds true during the free expansion and we arrive at the solution
\begin{equation}\label{eq:psi-long-time-0}
 \psi_\Mlabel{D}(t,\rr)\cong \frac{1}{\sqrt{\det\Lambda(t)}}
 \Me^{\frac{\Mi}{\hbar}\left[\frac{m}{2}\rr^\tp \diff{\Lambda}{t} \Lambda^{-1}\rr-\beta(t)\right]} \,
 \psi_\Mlabel{D}\left(0,\Lambda^{-1}(t) \,\rr\right),
\end{equation}
where we have used the definition \eqref{Def_A} of the symmetric matrix $A(t)$. 

In the long-time limit and for free expansion the adaptive matrix $\Lambda(t)$ grows linearly, Eq. \eqref{eq:long-time-lambda}, that is 
$\Lambda(t) \cong B\,t$ as $t \to \infty$ with the diagonal matrix $B = \op{diag}(b_1,\ldots,b_d)$. 
When we apply this limit to expression \eqref{eq:psi-long-time-0}, 
we obtain 
\begin{equation}\label{eq:psi-long-time-1}
  \psi_\Mlabel{D}(t,\rr)\cong \frac{1}{\sqrt{t^{d}\det B}} \,
  \Me^{\frac{\Mi}{\hbar} \left[\frac{m}{2t}\rr^2 - \beta(t)\right]} \,
  \psi_\Mlabel{D}\left(0,B^{-1}\frac{\rr}{t}\right).
\end{equation}
As a result, the spatial density distribution in the long time limit
\begin{equation}\label{eq:density-distribution-long-time}
  \left|\psi_\Mlabel{D}(t,\rr)\right|^2 \cong \frac{1}{t^d \det B} \,
  \left|\psi_\Mlabel{D}\left(0,B^{-1}\frac{\rr}{t}\right)\right|^2 
\end{equation}
is determined by the initial spatial density distribution and the time-dependent scaling. 
Obviously, Eq. \eqref{eq:density-distribution-long-time} is a direct consequence of the time-dependent TF approximation on which this derivation is based and stands in contrast to relation \eqref{eq:Schrodinger-result} derived for the Schr\"odinger equation. 

In the next step we obtain a similar relation for the momentum distribution of a free expanding BEC in the long-time limit. In order to do so, we insert Eq. \eqref{eq:psi-long-time-1} into Eq. \eqref{eq:Fourier-trafo-p-x} for the Fourier transform and arrive at the expression
\begin{equation}
 \label{eq:psi-tilde-long-time-1}
 \tilde\psi_\Mlabel{D}(t,\Mbf{p})\cong \frac{
 \Me^{-\frac{\Mi}{\hbar}\beta(t)-\frac{\Mi}{\hbar}\frac{\Mbf{p}^2 t}{2m} }}{(2\pi\hbar \,t)^{d/2}\sqrt{\det B}}
 \int_{\mathds{R}^d} \Me^{\frac{\Mi}{\hbar} \frac{m}{2t} \rr'^2}
 \psi_\Mlabel{D}\left(0, B^{-1}\frac{\rr'}{t} + B^{-1}\frac{\Mbf{p}}{m}\right)\Mdiff^d x',
\end{equation}
where we have introduced the new integration variable $\rr'\equiv \rr - \Mbf{p} t/m$. For long times $t\rightarrow \infty$, the argument of the initial wave function $\psi_\Mlabel{D}$ 
gets practically independent of $\rr'$, which allows to perform the remaining integration and to obtain
\begin{equation}
 \label{eq:psi-tilde-long-time-2}
 \tilde\psi_\Mlabel{D}(t,\Mbf{p})\cong 
 \frac{\Me^{-\frac{\Mi}{\hbar}\beta(t)-\frac{\Mi}{\hbar} \frac{\Mbf{p}^2 t}{2m} }}{(\Mi \,m)^{d/2}\, \sqrt{\det B}} 
 \psi_\Mlabel{D}\left(0, B^{-1}\frac{\Mbf{p}}{m}\right).
\end{equation}
Hence, in the case of the GP equation for free expansion, the momentum distribution in the long-time limit
\begin{equation}
 \label{eq:momentum-distribution-long-time}
 \left|\tilde\psi_\Mlabel{D}(t,\Mbf{p})\right|^2 \cong 
 \frac{1}{m^d\, \det B} \left|\psi_\Mlabel{D}\left(0, B^{-1}\frac{\Mbf{p}}{m}\right)\right|^2
\end{equation}
is given by the initial spatial density distribution, while for the Schr\"odinger equation it is vice versa, 
Eq. \eqref{eq:Schrodinger-result}. Moreover, according to Eq. \eqref{eq:momentum-distribution-long-time} 
the momentum distribution $|\tilde\psi_\Mlabel{D}(t,\Mbf{p})|^2$ of a free expanding BEC becomes time-independent in the long-time limit. 
However, we emphasize that for the derivation of Eq. \eqref{eq:momentum-distribution-long-time} 
we have required the validity of the time-dependent TF approximation during the free expansion. As a consequence, for a proper interpretation of time-of-flight pictures of BECs the interactions play an important role, especially deep within the TF regime. 

In addition, we point out that the methods presented in this section have been successfully applied by \cite{Roura2014} 
to study the influence of gravity gradients on the properties of an atom interferometer.

\section{Conclusion and Outlook}
In this chapter we have introduced a natural generalization of the scaling approach for time-dependent rotating traps, the so-called affine approach. 
In contrast to the hydrodynamical approach, we have carried out this generalization directly to the GP equation by first eliminating the center-of-mass motion and by subsequently applying a linear mapping to account for the main part of the remaining internal dynamics of a BEC. With the help of this affine approach we have established an efficient analytic description of the BEC dynamics based on two main assumptions: (i) the external potential in the vicinity of the BEC is at most quadratic and (ii) the time-dependent TF approximation holds true during the whole time evolution, which manifests itself in the fact that the density distribution of the condensate preserves its parabolic shape.

We have verified the accuracy of this approach by performing full numerical simulations of the GP equation for two scenarios: (i) a pure rotation of the trap, and (ii) a free expansion of an initially rotating BEC. In both cases we have found an excellent agreement of the numerical results with our approximate analytical solution for many different combinations of the interaction strength, the anisotropy factor and the final rotation rate of the trap. Moreover, the affine transformation can be successfully used to improve the performance of numerical simulations for a variety of experimentally relevant scenarios by solving the transformed GP equation rather than the original one.

By employing an additional transformation, we have introduced the canonical form of the matrix differential equation that governs the time evolution of the adaptive matrix and presented its corresponding Hamiltonian formalism. We have furthermore analyzed the relation between the constants of motion of this matrix differential equation and the constants of motion of the GP equation in connection with the conservation of the total energy and the angular momentum of a BEC. For both cases, we have derived explicit expressions valid within the TF regime.

In agreement with our affine approach, the {\it spatial} density distribution of a freely expanding BEC in the long-time limit {\it does not} display the initial {\it momentum} distribution in general, unlike a freely expanding atomic cloud that evolves according to the Schr\"odinger equation. Instead, it resembles the affinely transformed initial spatial density distribution. However, there exists a converse relation between the spatial and momentum distributions for a BEC within the TF regime, that is the {\it momentum} distribution of a free expanding BEC in the long-time limit is indeed given by its initial {\it spatial} density distribution. These relations are a direct consequence of the nonlinearity of the GP equation and stand in contrast to the well-known ballistic expansion that wave functions governed by the free Schr\"odinger equation show. Hence, in order to properly interpret time-of-flight pictures of BECs, the mean-field interaction has to be taken into account. This observation is of special importance for the TF regime where the interactions play a dominant role. 

In summary, scaling solutions like the affine approach presented in this chapter are valuable tools for studying the dynamics of strongly interacting quantum gases. Not only do they provide efficient ways to accurately describe their time evolution, but they also allow to unravel and separate the different layers of their dynamics. Similar approaches have been successfully used in a multitude of interesting physical phenomena, such as (i) the Gaussian wave packet, which keeps its form in an at most harmonic potential, (ii) the Airy wave packet \cite{Berry1979,Kajari2010} and (iii) the Bessel beam \cite{Durnin1987,DurninEberly1987}, which freely propagate without dispersion. 

We emphasize that there are a few scenarios in which one of the two above-mentioned assumptions for our efficient analytic description of the BEC dynamics is not fulfilled. Indeed, the external potential could be of a form incompatible with the harmonic expansion, Eq. \eqref{eq:potential}, as is the case for (i) realistic trapping potentials created by optical dipole traps, optical lattices and magnetic chip traps, where the anharmonic terms play a crucial role especially in the context of delta-kick collimation techniques \cite{ammann-1997,Zeller2016}, and (ii) trapping potentials which are not differentiable due to sharp corners like linear potential wells (the irrigation canal) \cite{Gallas1995,Bestle1995} or gravitational traps \cite{Aminoff1993,Nesvizhevsky2002}. 
Moreover, the TF approximation is well-known to break down when the interaction strength approaches very small values, which can be caused by a low number of atoms, or a decreasing coupling constant. In order to accurately describe BECs in this regime one can either expand the theoretical model beyond the TF approximation \cite{Jamison2011} or resort to numerical methods based on the affinely transformed GP equation.
For all these scenarios the affine approach can still offer valuable insights by dedicated numerical simulations that take advantage of the affine transformation to substantially reduce the computational costs of the simulations \cite{PhD_ME}. 

Finally, it is noteworthy that the affine approach can even be applied to multi-species BECs to study their expansion dynamics both from an analytical and numerical perspective. However, a detailed discussion of this method is beyond the scope of this chapter and will be addressed in a future publication \cite{Meister2017}.

\section*{Acknowledgments}
We would like to thank H.~Ahlers, M.~Buser, A.~Friedrich, N.~Gaaloul, E.~Giese, N.~Harshman, W.~Herr, J.~Jenewein, S.~Kleinert, M.~Krutzik, W.~Lewoczko-Adamczyk, A.~Roura, E.~Sadurni, S.~T.~Seidel, E.~M.~Rasel, C.~Ufrecht, V.~Yakovlev, W.~Zeller and T.~van~Zoest for many fruitful discussions and helpful suggestions. M.~A.~E. thanks the Alexander von Humboldt Stiftung and W.~P.~S. is grateful to the Hagler Institute for Advanced Study at Texas A\&M University for a Faculty Fellowship. This project was generously supported by the German Space Agency (DLR) with funds provided by the Federal Ministry for Economic Affairs and Energy (BMWi) under the grant numbers 50WM0346, 50WM0837, 50WM1136 and 50WM1556.

\appendix

\section{Affine transformation}
\label{AppAffineTrafo}

In this appendix we outline the derivation of the affinely transformed GP equation~\eqref{ScaledGP}. This is done by first
eliminating the center-of-mass motion of the condensate and second, by performing
a linear transformation of the coordinates together with a unitary
transformation of the wave function. Moreover, we show how to determine the 
integrated density distributions of a BEC within the TF regime.

\subsection{Center-of-mass motion\label{App_COM}}
\label{AppCenterofmass}

Our starting point is the GP equation \eref{GPequation} 
\begin{equation}
 \label{GPequationTrap}
\fl
\Mi\hbar\,\pdiff{\psi}{t}= \left[-\frac{\hbar^2}{2m}\bnabla^2_{\!\rr}+ V(t,\brho) - \bgg (\rr\!-\!\brho)
+\frac{m}{2}(\rr-\brho)^{\!\tp}\Omega^2\,(\rr\!-\!\brho) + g\,|\psi|^2\right]\psi
\end{equation}
with the explicit time-dependent harmonic potential given by Eq. \eref{eq:potential}.
We then rewrite the transformation~\eqref{DefpsiD} by making use of the displacement operator in 
position representation
\begin{equation}
 \label{D-appendix}
 \hat D(\calR,\calP)\equiv \Me^{\frac{\Mi}{\hbar}(\calP\,\rr-\calR\,\bpp)},
\end{equation} 
and obtain 
\begin{equation}
 \label{DefpsiD-appendix}
 \psi(t,\rr) = e^{\frac{i}{\hbar}\calS_1(t)}\hat D(\calR,\calP)\psi_\Mlabel D(t,\rr)\,.
\end{equation} 
The displacement operator satisfies the relations 
\begin{eqnarray} 
&&\hat
D(\calR,\calP)\,\psi_\Mlabel D(t,\rr)=\Me^{-\frac{\Mi}{2\hbar}\calR\calP}\Me^{\frac{\Mi}{\hbar}\calP\,\rr}
\psi_\Mlabel D(t,\rr-\calR) \,,
\label{displacewavefunction}\\
&&\hat D^\dagger(\calR,\calP)\,f(\rr,\bpp)\,\hat D(\calR,\calP)=f(\rr+\calR,\bpp+\calP)
\label{propertydisplace}
\end{eqnarray} 
for any smooth wave function $\psi_\Mlabel D(t,\rr)$ and phase-space operator $f(\rr,\bpp)$.
Taking advantage of Eqs.~\eqref{DefpsiD-appendix} - \eqref{propertydisplace}, 
as well as the fact that
\begin{equation*}
\hat D^\dagger\,|\psi(t,\rr)|^2\,\hat D=
\hat D^\dagger\,|\psi_\Mlabel D(t,\rr-\calR)|^2\,\hat D=|\psi_\Mlabel D(t,\rr)|^2\,,
\end{equation*} 
we arrive at the transformed GP equation
\begin{eqnarray}
\label{trafoGP}
\fl
\Mi\hbar\,\pdiff{\psi_\Mlabel D}{t}
=\left\{\frac{\bpp^2}{2m}
    +\frac{m}{2}\rr^{\!\tp}\Omega^2\,\rr+g\,|\psi_\Mlabel D|^2\right.
    +\left[-\bgg\!+\!m\Omega^2\left(\calR\!-\!\brho\right)\!+\!\diff{\calP}{t}\right]\rr
    +\left[\frac{\calP}{m}\!-\!\diff{\calR}{t}\right]\bpp \nonumber \\ 
  +\diff{\calS_1}{t}+\diff{}{t}\left(\frac{\calR\calP}{2}\right)
  -\diff{\calR}{t}\,\calP +\frac{\calP^2}{2m}+V(t,\brho) -\bgg (\calR\!-\!\brho)\nonumber \\
  +\left. \frac{m}{2} (\calR\!-\!\brho)^{\!\tp}\Omega^2\left(\calR\!-\!\brho\right)\right\}\psi_\Mlabel D
\end{eqnarray}
for the wave function $\psi_\Mlabel D$.

In order to eliminate all terms in Eq. \eqref{trafoGP} that depend linearly on $\rr$ and $\bpp$, as well as arbitrarily on~$\brho$, we require the parameters $\calR(t)$ and $\calP(t)$ to satisfy the classical equations of motion~\eref{eq:equationsOfMotion}
and the phase $\calS_1(t)$ to be a solution of the first order differential equation
\begin{equation}
 \label{DGLcalS}
 \diff{\calS_1}{t}=\calL(\calR,\dot{\calR},t)-\frac{1}{2}\diff{}{t}\left(\calR\calP\right)\,.
\end{equation}
Here, we have used the first equation in Eq.~\eqref{eq:equationsOfMotion} and the definition 
of the Lagrange function $\calL(\calR,\dot{\calR},t)$, Eq. \eqref{eq:Lagrangian}. As a result, we have simplified the
transformed GP equation~\eqref{trafoGP} to Eq.~\eqref{GPequationD}.

\subsection{Linear transformation\label{App_Lambda}}

We continue the derivation of the affinely transformed GP equation 
by recalling Eq.~\eqref{lineartrafo} for the linear transformation of the spatial coordinates and the substitution of the  time variable. This mapping implies that the corresponding partial derivatives in Eq. \eqref{GPequationD} transform as
\begin{eqnarray}
 \label{partialderivatives}
\pdiff{}{t}& =& \pdiff{}{\tau}- 
\left(\Lambda^{-1}(\tau)\,\pdiff{\Lambda}{\tau}\,\bzeta\right)^{\!\!\tp}\bnabla_{\bzeta}\,, \nonumber\\
\bnabla_\rr & = & \Lambda^{-\tp\!}(\tau)\,\bnabla_{\bzeta}\,.
\end{eqnarray}
As mentioned in subsection~\ref{SecLinearTrafo}, the linear mapping~\eqref{lineartrafo} is accompanied by the 
unitary transformation of the wave function~\eqref{DefpsiLambda} which guarantees that the affinely 
transformed wave function $\psi_\Mlabel \Lambda(\tau,\bzeta)$ is again normalized according to 
Eq.~\eqref{eq:Normalization}.
Thus, by using the Eqs.~\eqref{lineartrafo}, \eqref{DefpsiLambda} and \eqref{partialderivatives}, as well as the identity
\begin{equation}
  \diff{}{\tau}\big(\det \Lambda\big)
  = \det\Lambda\cdot\tr{\Lambda^{-1}\,\diff{\Lambda}{\tau}}\,,
\end{equation}
we arrive at the GP equation for the affinely transformed wave function $\psi_\Mlabel\Lambda(\tau,\bzeta)$
\begin{eqnarray}
  \Mi\hbar\,\pdiff{\psi_\Mlabel\Lambda}{t}
  &=& \left(- \frac{\hbar^2}{2m}
      \bnabla^\tp_{\bzeta}\Lambda^{-1}\!\Lambda^{-\tp}\bnabla_{\bzeta}
    + \frac{g}{\det \Lambda}|\psi_\Mlabel\Lambda|^2
    \right)\psi_\Mlabel\Lambda
    \nonumber\\
  && + \Mi\hbar \,
    \psi_\Mlabel\Lambda\left[\tr{\frac{\Lambda^{-1}}{2}
    \left(\diff{\Lambda}{\tau}
    - \frac{2}{m} \Lambda^{-\tp} {A}\right)}
    + \frac{\Mi}{\hbar}\diff{\beta}{\tau}\right]\nonumber\\
  && + \psi_\Mlabel\Lambda\,\bzeta^\tp\hspace{-0.5ex}
    \left[\diff{{A}}{\tau}
    - 2 {A}\,\Lambda^{-1}\hspace{-0.5ex}
    \left(\diff{\Lambda}{\tau}-\frac{1}{m}\Lambda^{-\tp} {A}\right)
    \hspace{-0.5ex}
    + \frac{m}{2}\Lambda^\tp\Omega^2\Lambda
    \right]\hspace{-0.4ex} \bzeta
    \nonumber\\
  && + \Mi\hbar\,\big(\bnabla_{\bzeta}\psi_\Mlabel\Lambda\big)^{\!\tp}
  \Lambda^{-1}
  \left[\diff{\Lambda}{\tau}-\frac{2}{m} \Lambda^{-\tp} {A}\right]\bzeta\,.
\label{ScaledGPbefore}
\end{eqnarray}

Now we first choose $A(\tau)$ to be given by Eq. \eqref{Def_A} 
which serves the purpose of eliminating the last term in Eq. \eqref{ScaledGPbefore} as well as the part that 
contains the trace in the second term on the right-hand side.
Next we insert $\beta(\tau)$ as defined by Eq. \eqref{Def_beta}. 
In order to simplify the third term on the right-hand side of Eq. \eqref{ScaledGPbefore}, we therein
replace $A(\tau)$ according to Eq. \eqref{Def_A} and require the adaptive matrix $\Lambda(\tau)$ to satisfy the matrix differential equation \eqref{MatrixDGL}. Finally, by taking advantage of the irrotationality condition~\eqref{eq:IrrotCondition} whose validity is a direct  
consequence of Eqs.~\eqref{MatrixDGL} and~\eqref{InitialMatrix} as discussed in the next section, we achieve our goal to 
simplify Eq.~\eqref{ScaledGPbefore} to the affinely transformed GP equation~\eqref{ScaledGP}.

\subsection{Irrotationality condition}\label{sec:app:irrotationality-condition}
Within our approach the adaptive matrix $\Lambda(\tau)$ automatically satisfies 
the irrotationality condition, Eq. \eqref{eq:IrrotCondition}, since the time derivative of the auxiliary matrix function
\begin{equation*}
  Z(\tau)
  = \Lambda^{\tp\!}(\tau)\,\diff{\Lambda}{\tau}
  - \diff{\Lambda^\tp}{\tau}\,\Lambda(\tau) 
\end{equation*} 
vanishes for all ${\tau\geq 0}$ provided $\Lambda(\tau)$ is a solution of the matrix differential equation~\eqref{MatrixDGL} 
with the initial conditions~\eqref{InitialMatrix}. For this reason, the antisymmetric matrix $Z(\tau) = Z(0) = 0$ 
represents $d(d-1)/2$ constants of motion of the second order matrix differential equation~\eqref{MatrixDGL} and 
the matrix~$A(\tau)$, Eq.~\eqref{Def_A}, is indeed symmetric.

However, as pointed out by \cite{storey00}, the irrotationality condition \eqref{eq:IrrotCondition} can likewise be derived by using the notion of classical particle trajectories that constitue a velocity field $\Mbf{v}(t,\rr)$, which is irrotational, namely $\bnabla\times\Mbf{v}(t,\rr)  = 0$ at any point in space. We briefly sketch this derivation within our approach. Let $\rr = \rr(\rr_0,t)$ be the classical trajectory of a particle within a BEC, which depends on the initial position $\rr_0$ and the time $t$. As we have shown in section \ref{sec:effective_description} a BEC does not show any time evolution in the adapted coordinates if the time-dependent TF approximation holds true. Hence, the trajectory in adapted coordinates is given by $\bzeta(t) = \bzeta_0 = \rr_0$, keeping in mind that the original and the adapted coordinates coincide at $t=0$. Since the transformation between the two coordinate systems is defined by the linear mapping $\bzeta = \Lambda^{-1}(t)\left[\rr-\calR(t)\right]$ with the 
center-of-mass position $\calR(t)$ and the adaptive matrix $\Lambda(t)$, the classical trajectory is given by
\begin{equation}
 \rr_0=\Lambda^{-1}(t)\left[\rr-\calR(t)\right]
\end{equation}
or rather
\begin{equation}
 \rr(\rr_0,t) = \calR(t) + \Lambda(t) \,\rr_0 \;.
\end{equation}
Thus, the velocity field
\begin{equation}
 \boldsymbol{v}(\rr,t) \equiv \diff{\rr}{t} =\diff{\calR}{t} + \diff{\Lambda}{t}\,\Lambda^{-1}\,(\rr - \calR)
\end{equation}
is irrotational, that is $\bnabla\times\Mbf{v}(t,\rr)  = 0$, 
if the matrix $\left(\Mdiff\Lambda/\Mdiff t\right)\Lambda^{-1}$ is symmetric
\begin{equation}
 \Lambda^{-\tp}\,\diff{\Lambda^\tp}{t} = \diff{\Lambda}{t}\,\Lambda^{-1},
\end{equation}
which is equivalent to the irrotationality condition \eqref{eq:IrrotCondition}.

\subsection{Integration of the density distribution of a Bose-Einstein condensate within the Thomas-Fermi regime}
\label{sec:appDensityIntegration}

In this appendix, we discuss the integration of the density distribution 
$\abs{\psi_\Mlabel{TF}(t,\rr)}^2$ of a \mbox{$d$-dimensional}  BEC within the TF 
regime over $n$ dimensions with $n<d$. 
We denote by \(\rr_{[2]}\in\Reals^n\) the vector consisting of the components of
\(\rr\in\Reals^d\) over which the integration is performed. Accordingly, the vector \(\rr_{[1]}\in\Reals^{d-n}\) 
is composed of the remaining coordinates. Thus, the integrated density distribution is defined as
\begin{equation}
  \label{eq:integratedDensityDefinition}
  n_\Mlabel{TF}(t,\rr_{[1]})
 = \int_{\Reals^n} 
    \abs{\psi_\Mlabel{TF}(t,\rr)}^2\,\Mdiff^{n\!} x_{[2]}\,.
\end{equation}

Without loss of generality, we can rearrange the coordinates in the original density distribution, Eq. \eqref{eq:TimeEvolTF_Density},
such that \(\rr^\tp=(\rr_{[1]}^\tp,\rr_{[2]}^\tp)\), which furthermore translates into ${\calR^\tp=(\calR_{[1]}^\tp,\!\calR_{[2]}^\tp)}$ for the center-of-mass position and 
\begin{equation*}
  \Sigma
  = \left(\begin{array}{*{2}{c}}
    \Sigma_{[11]} & \Sigma_{[12]} \\
    \Sigma_{[12]}^{\tp} & \Sigma_{[22]}
  \end{array}\right)
\end{equation*}
for the TF matrix defined by Eq. \eqref{eq:TF_Matrix}.
By using the decomposition of the matrix $\Sigma$ into the Schur complements \cite{HornJohnson1990}
\begin{equation}
  \fl
  \Sigma
  = \left(
    \begin{array}{*{2}{c}}
      \Unit_{d-n}
      & \hspace{-1mm}\Zero \\
      \Sigma_{[12]}^{\,\tp}\Sigma_{[11]}^{-1}
      & \Unit_{n} \\
    \end{array}%
   \right)
    \left(
    \begin{array}{*{2}{c}}
     \Sigma_{[11]}
      & \Zero \\
      \Zero
      & \Sigma_{[22]} \!-\!
      \Sigma_{[12]}^{\,\tp}\Sigma_{[11]}^{-1}
      \Sigma_{[12]}
      \\ \end{array}%
   \right)%
   \left(
    \begin{array}{*{2}{c}}
      \Unit_{d-n}
      & \hspace{-1mm}\Zero \\
      \Sigma_{[12]}^{\,\tp}\Sigma_{[11]}^{-1}
      & \Unit_{n} \\
    \end{array}%
   \right)^{\!\!\!\tp}
   \label{eq:Schur_Complement}\,,
\end{equation}
the inverse of the TF matrix can be written as
\begin{equation*}
  \fl
  \Sigma^{-1}
  \!=\! \left(
    \begin{array}{*{2}{c}}
      \Unit_{d-n}
      & \Zero \\
      -\Sigma_{[12]}^{\,\tp}\Sigma_{[11]}^{-1}
      & \Unit_{n} \\
    \end{array}%
    \right)^{\!\!\tp}
    \left(%
    \begin{array}{*{2}{c}}
      \Sigma_{[11]}^{-1}
      & \Zero \\
      \Zero
      & \big(\Sigma_{[22]}\!-\!
      \Sigma_{[12]}^{\,\tp\phantom{-}}\Sigma_{[11]}^{-1}
      \Sigma_{[12]}\big)^{\!-1}
      \\ \end{array}%
    \right)\!\!
    \left(
    \begin{array}{*{2}{c}}
      \Unit_{d-n}
      & \Zero \\
      -\Sigma_{[12]}^{\,\tp}\Sigma_{[11]}^{-1}
      & \Unit_{n} \\
    \end{array}%
    \right)\,.
\end{equation*}
By inserting this expression into Eq.~\eqref{eq:TimeEvolTF_Density}, the quadratic 
form $(\rr-\calR)^{\tp}\Sigma^{-1}(\rr-\calR)$ splits  
into two terms: $(i)$ a term involving only 
the free coordinates $\rr_{[1]}$ and $(ii)$ a term depending on both, the coordinates $\rr_{[2]}$ over which the integration is performed and the free coordinates $\rr_{[1]}$. In view of the integration in Eq.~\eqref{eq:integratedDensityDefinition}, 
we perform a change of the integration variables from $\rr_{[2]}$ to $\Mbf{y}$ via
\begin{equation*}
  \Mbf{y}
  = \sqrt{\frac{m}{2}}\,\Big(\Sigma_{[22]}
    -\Sigma_{[12]}^{\,\tp}\Sigma_{[11]}^{-1}
    \Sigma_{[12]}\Big)^{-\frac{1}{2}}
    \left[\rr_{[2]}\!-\!\calR_{[2]}
      \!-\!\Sigma_{[12]}^{\,\tp}\Sigma_{[11]}^{-1}
      \left(\rr_{[1]}\!-\!\calR_{[1]}\right)\!
    \right]\,.
\end{equation*}
Note, that the inverse square root of the matrix $(\Sigma_{[22]}
    -\Sigma_{[12]}^{\,\tp}\Sigma_{[11]}^{-1}
    \Sigma_{[12]})$ is defined in terms of its spectral decomposition.
As a result, we obtain for the integrated density distribution~\eqref{eq:integratedDensityDefinition}
\begin{equation}
\label{eq:Integrated_Density_2}
\fl
  n_\Mlabel{TF}(t,\rr_{[1]})
  = \frac{1}{g\det\Lambda}\sqrt{\frac{2^n}{m^n}
  \det\!\Big(\Sigma_{[22]}
    -\Sigma_{[12]}^{\,\tp\phantom{-}}\Sigma_{[11]}^{-1}
    \Sigma_{[12]}\Big)}
    \int_{\Reals^n}\!\!
    \left\{R^2(\rr_{[1]})-\Mbf{y}^2\right\}_+\Mdiff^n y,
\end{equation}
where we have introduced the auxiliary function
\begin{equation*}
  R(\rr_{[1]})\equiv\left\{
    \mu_\Mlabel{TF}
    -\frac{m}{2}\left(\rr_{[1]}\!-\!\calR_{[1]}\right)^\tp
      \Sigma_{[11]}^{-1}\left(\rr_{[1]}\!-\!\calR_{[1]}\right)\right\}_+^{\frac{1}{2}}\,.
\end{equation*}
We perform the integration in Eq.~\eqref{eq:Integrated_Density_2} with the help of 
the $n$-dimensional spherical coordinates
\begin{equation}
 \label{n-integration-appendix}
   \fl
   \int_{\Reals^n}\!\!
    \left\{R^2-\Mbf{y}^2\right\}_+\Mdiff^n y=
    \int_{S_{n-1}}\!\!\!\!\Mdiff^{n-1}\!\Omega\int_{0}^{R} (R^2-r^2)\,r^{n-1}\,\Mdiff r =
  \frac{4\pi^{\frac{n}{2}}}{n(n+2)\Gamma\left(\frac{n}{2}\right)}R^{n+2},
\end{equation}
where $S_{n-1}$ denotes the $(n-1)$-dimensional spherical surface. 
Moreover, by applying the determinant on both sides of the decomposition~\eqref{eq:Schur_Complement},
we find with the help of Eq.~\eqref{eq:TF_Matrix} the useful relation
\begin{equation*}
  \det\Big(\Sigma_{[22]\!}
    -\Sigma_{[12]\!}^{\,\tp}\,\Sigma_{[11]\!}^{-1}\,
    \Sigma_{[12]}\Big)
  = \frac{\det\Sigma}{\det\Sigma_{[11]}}
  = \frac{(\det\Lambda)^2}{\det\Omega^{2\!}(0)\det\Sigma_{[11]}}\,.
\end{equation*}
Taking advantage of the latter, the integrated density distribution Eq. \eqref{eq:Integrated_Density_2} finally reads
\begin{equation*}
  \fl
  \textstyle{
  n_\Mlabel{TF}(t,\rr_{[1]})
  = \frac{1}
    {\GammaF\left(\frac{n}{2}+2\right)\,g}
    \sqrt{\frac{(2\pi/m)^n}{\det\Omega^{2\!}(0)\det\Sigma_{[11]\!}}} 
    \left\{\mu_\Mlabel{TF}
      - \frac{m}{2}\left(\rr_{[1]}\!-\!\calR_{[1]}\right)^{\!\tp}\Sigma_{[11]}^{-1}\left(\rr_{[1]}\!-\!\calR_{[1]}\right)
    \right\}_+^{\frac{n}{2}+1}}.
\end{equation*}
For $n=1$ and $n=2$, this result reduces to Eqs. \eqref{eq:integratedDensit2d} and \eqref{eq:integratedDensit1d}, respectively.

\section{Energy and angular momentum of a Bose-Einstein condensate}\label{sec:app:constants_of_motion}

This appendix presents the important steps to determine the different contributions 
to the total energy of a BEC and its angular momentum within the time-dependent TF approximation.

\subsection{Energy terms in the time-dependent Thomas-Fermi approximation}
\label{sec:appEnergy}

We now evaluate the energy terms, Eqs. \eqref{eq:kinEnergy:def} - \eqref{eq:intEnergy:def},
by using the approximate wave function, Eq. \eqref{eq:general:EffectiveTimeEvolTF}, 
valid in the time-dependent TF approximation. 
With the help of Eq. \eqref{eq:Ehrenfest_substitutions} 
we can rewrite the expectation values in Eqs. \eqref{eq:kinEnergy:def} - \eqref{eq:intEnergy:def} 
in terms of the expectation value with respect to the initial state $\psi(0,\rr)$.
Thus, for the kinetic energy, Eq. \eqref{eq:kinEnergy:def}, we obtain
\begin{eqnarray}
\fl 
\frac{E_\Mlabel{kin} (t)}{N} 
 = \frac{1}{2m} \left[\calP(t) - \Lambda^{-\tp}(t)\, \calP(0) \right]^2 
 + \frac{1}{m} \left[\calP(t) - \Lambda^{-\tp}(t)\calP(0)\right]^\tp \Lambda^{-\tp}(t) \ave{\bpp}_{\psi_0}    \nonumber \\
\fl\qquad\qquad 
 + \frac{1}{2m} \ave{\bpp^\tp\Lambda^{-1}(t)\Lambda^{-\tp}(t)\,\bpp }_{\psi_0}
 + \frac{m}{2} \ave{\left[\hat\rr - \calR(0)\right]^\tp \diff{\Lambda^\tp}{t}\diff{\Lambda}{t} 
\left[\hat\rr - \calR(0)\right]}_{\psi_0} \nonumber \\
\fl\qquad\qquad
 + \left[\calP(t) - \Lambda^{-\tp}(t)\calP(0)\right]^\tp \diff{\Lambda}{t} 
\left[\ave{\hat\rr}_{\psi_0} - \calR(0)\right] \nonumber \\
\fl\qquad\qquad
 + \mathrm{Re}\left\lbrace \ave{\left[\hat\rr - \calR(0)\right]^\tp \diff{\Lambda^\tp}{t} \Lambda^{-\tp}(t)\,\bpp}_{\psi_0} \right\rbrace .
 \label{eq:kinEnergy:1step}
\end{eqnarray}
In the same manner the potential energy given by Eqs. \eqref{eq:potEnergy:def} and \eqref{eq:potential} reads
\begin{eqnarray}
\fl \frac{E_\Mlabel{pot} (t)}{N} = V(t,\brho(t)) 
+ \frac{m}{2}\ave{\left[\hat\rr - \calR(0)\right]^\tp\Lambda^\tp(t)\Omega^2(t)\Lambda(t)
\left[\hat\rr - \calR(0)\right]}_{\psi_0} \nonumber \\
\fl\qquad\qquad - \bgg(t)\Lambda(t)\left[\ave{\hat\rr}_{\psi_0} - \calR(0)\right]
- \bgg(t)\left[\calR(t) - \brho(t)\right] \nonumber \\
\fl\qquad\qquad + m \left[\calR(t) - \brho(t)\right]^\tp \Omega^2(t)\Lambda(t)
\left[\ave{\hat\rr}_{\psi_0}-\calR(0)\right] \nonumber \\
\fl\qquad\qquad + \frac{m}{2} \left[\calR(t) - \brho(t)\right]^\tp \Omega^2(t)\left[\calR(t)-\brho(t)\right].
\label{eq:potEnergy:1step}
\end{eqnarray}
Likewise, the dependence on time of the interaction energy Eq. \eqref{eq:intEnergy:def} is given by
\begin{equation}\label{eq:intEnergy:1step}
 \frac{E_\Mlabel{int} (t)}{N} = \frac{E_\Mlabel{int}(0)}{N\det\Lambda(t)} \,.
\end{equation}

We can further simplify Eqs. \eqref{eq:kinEnergy:1step} and \eqref{eq:potEnergy:1step} for the kinetic and the potential energies
by taking into account the initial conditions \eqref{eq:RP_initial_conditions} and 
by using the identity
\begin{equation}
\ave{\hat\Mbf{f}^\tp M \,\hat\Mbf{f} } = \sum\limits_{j,k} M_{jk} \ave{\hat{f}_j\, \hat{f}_k} = 
\mathrm{Tr}\left[M \ave{\hat\Mbf{f} \otimes \hat\Mbf{f}^\tp} \right],
\end{equation}
which is valid for any symmetric matrix $M$. Hence, Eqs. \eqref{eq:kinEnergy:1step} and \eqref{eq:potEnergy:1step} reduce to
\begin{eqnarray}
\fl \frac{E_\Mlabel{kin} (t)}{N} 
 = \frac{\calP^2(t)}{2m}
 + \frac{1}{2m} \,\mathrm{Tr} \left[\Lambda^{-1}(t)\,\Lambda^{-\tp}(t) \,\ave{\bpp \otimes \bpp^\tp}_{\psi_0} \right] 
 - \frac{1}{2m} \left[\Lambda^{-\tp}(t)\, \calP(0) \right]^2   \nonumber \\
\fl\qquad\qquad
 + \frac{m}{2} \,\mathrm{Tr} \left\{\diff{\Lambda^\tp}{t} \diff{\Lambda}{t}
\ave{\left[\hat\rr - \calR(0)\right] \otimes \left[\hat\rr - \calR(0)\right]^\tp}_{\psi_0} \right\} \nonumber \\
\fl\qquad\qquad 
 + \mathrm{Re}\left\lbrace \ave{\left[\hat\rr - \calR(0)\right]^\tp \diff{\Lambda^\tp}{t} \Lambda^{-\tp}(t)\,\bpp}_{\psi_0} \right\rbrace \label{eq:kinEnergy:2step}
\end{eqnarray}
and
\begin{eqnarray}
\fl \frac{E_\Mlabel{pot} (t)}{N} = V(t,\calR(t))+ \frac{m}{2} \,\mathrm{Tr} 
\left\{\Lambda^\tp(t)\Omega^2(t)\Lambda(t)
\ave{\left[\hat\rr-\calR(0)\right] \otimes \left[\hat\rr-\calR(0)\right]^\tp}_{\psi_0}\right\} \,,
\label{eq:potEnergy:2step}
\end{eqnarray}
respectively, where $V(t,\calR(t))$ denotes the harmonic potential \eqref{eq:potential}.

In addition, by making use of the explicit form of the initial wave function, Eq. \eqref{eq:initial:wave:function:inertial:frame}, we can evaluate the second and last term of Eq. \eqref{eq:kinEnergy:2step}. We start by inserting Eq. \eqref{eq:initial:wave:function:inertial:frame} into the expectation value of the second term of Eq. \eqref{eq:kinEnergy:2step} and arrive at the relation
\begin{equation}
 \fl\ave{\bpp \otimes \bpp^\tp}_{\psi_0} = \ave{\bpp \otimes \bpp^\tp}_{|\psi_0|} + \,\calP(0) \otimes \calP^\tp(0) \,+\, \ave{\bpp}_{|\psi_0|} \otimes \calP^\tp(0) \,+\, \calP(0) \otimes \ave{\bpp}^\tp_{|\psi_0|} \,,
\end{equation}
where the expectation values are now evaluated with respect to the absolute value of the initial state. 
Since the BEC is initially at rest in the comoving frame of the trap, we obtain $\ave{\bpp}_{|\psi_0|} = 0$ and thus the second term of Eq. \eqref{eq:kinEnergy:2step} fully reads
\begin{eqnarray}
 \frac{1}{2m} \,\mathrm{Tr} \left[\Lambda^{-1}(t)\,\Lambda^{-\tp}(t) \,\ave{\bpp \otimes \bpp^\tp}_{\psi_0} \right] \nonumber \\
\qquad 
 = \frac{1}{2m} \,\mathrm{Tr} \left[\Lambda^{-1}(t)\,\Lambda^{-\tp}(t) \,\ave{\bpp \otimes \bpp^\tp}_{|\psi_0|} \right]
 + \frac{1}{2m} \left[\Lambda^{-\tp}(t)\, \calP(0) \right]^2 \,. \label{eq:app:quantum:pressure:step0}
\end{eqnarray}
Next, we apply  Eq. \eqref{eq:initial:wave:function:inertial:frame}, to the last term of Eq. \eqref{eq:kinEnergy:2step} and arrive at the expression
\begin{eqnarray}\label{eq:app:energy:real:part:term}
 \mathrm{Re}\left\lbrace \ave{\left[\hat\rr - \calR(0)\right]^\tp \diff{\Lambda^\tp}{t} \Lambda^{-\tp}(t)\,\bpp}_{\psi_0} \right\rbrace \nonumber \\
\qquad
  = \left[\ave{\hat\rr}_{\psi_0} - \calR(0)\right]^\tp \diff{\Lambda^\tp}{t} \Lambda^{-\tp}(t)\,\calP(0) = 0 \,,
\end{eqnarray}
which vanishes if we take into account the initial conditions, Eq. \eqref{eq:RP_initial_conditions}, for the center-of-mass position $\calR(0)$ of the BEC. 
Thus, by inserting Eqs. \eqref{eq:app:quantum:pressure:step0} and \eqref{eq:app:energy:real:part:term} into Eq. \eqref{eq:kinEnergy:2step}, the kinetic energy is finally given by
\begin{eqnarray}
\frac{E_\Mlabel{kin} (t)}{N} 
 = \frac{\calP^2(t)}{2m}
 + \frac{1}{2m} \,\mathrm{Tr} \left[\Lambda^{-1}(t)\,\Lambda^{-\tp}(t) \,\ave{\bpp \otimes \bpp^\tp}_{|\psi_0|} \right]  \nonumber \\
\qquad\qquad
 + \frac{m}{2} \,\mathrm{Tr} \left\{\diff{\Lambda^\tp}{t} \diff{\Lambda}{t}
\ave{\left[\hat\rr - \calR(0)\right] \otimes \left[\hat\rr - \calR(0)\right]^\tp}_{\psi_0} \right\} \,. \label{eq:kinEnergy:3step}
\end{eqnarray}

In conclusion, by summing up the contributions given by Eqs. \eqref{eq:intEnergy:1step}, \eqref{eq:potEnergy:2step} and \eqref{eq:kinEnergy:3step}, we arrive at Eq. \eqref{eq:totalEnergy:all:terms} for the energy per particle of a BEC within the time-dependent TF approximation.

\subsection{Energy terms for the Thomas-Fermi ground state}
\label{sec:appEnergy:TF:groundState}

In this appendix we use the TF ground state $\phi_\Mlabel{TF}(\rr)$, Eq. \eqref{eq:initialTFwavefunc}, to derive explicit expressions for the energy of a BEC given by Eq. \eqref{eq:totalEnergy:all:terms}. 
Since we have neglected the quantum pressure term in the derivation of the TF ground state, Eq. \eqref{eq:initialTFwavefunc}, we consequently neglect it also for the calculation of the energy within the TF approximation, that is
\begin{eqnarray}\label{eq:app:quantum:pressure:step2}
 \ave{\bpp \otimes \bpp^\tp}_{|\phi_\Mlabel{TF}|} \approx 0 \,.
\end{eqnarray}

Next we use Eq. \eqref{eq:initialTFwavefunc}, the transformation 
$\bzeta = \rr - \calR(0)$, and the relation \eqref{eq:chemicalPotentialTF} to rewrite the expectation value 
\begin{equation*}
\fl \ave{\big[\hat\rr - \calR(0)\big] \otimes \big[\hat\rr - \calR(0)\big]^\tp}_{\phi_\Mlabel{TF}} = 
\frac{1}{Ng} \int\limits_{\Reals^d}\bzeta \otimes \bzeta^\tp 
\left[\mu_\Mlabel{TF} - \frac{m}{2}\, \bzeta^\tp \Omega^2(0) \,\bzeta \right]_+\; \Mdiff^d \xi
\end{equation*}
in the form
\begin{equation}\label{eq:TFexpect:square:position:step2}
 \ave{\big[\hat\rr - \calR(0)\big] \otimes \big[\hat\rr - \calR(0)\big]^\tp}_{\phi_\Mlabel{TF}} = 
 \frac{2}{d+4} \,\frac{\mu_\Mlabel{TF}}{m} \,\Omega^{-2}(0)\,,
\end{equation}
which is valid in $d$ dimensions. Similarly, by applying Eq. \eqref{eq:initialTFwavefunc} to Eq. \eqref{eq:intEnergy:def} we obtain
\begin{equation}\label{eq:TF:initial:intEnergy}
 E_\Mlabel{int}(0) = \frac{2}{d+4}\, N\, \mu_\Mlabel{TF}\,.
\end{equation}

Finally, when we insert Eqs. \eqref{eq:app:quantum:pressure:step2} - \eqref{eq:TF:initial:intEnergy} into Eq. \eqref{eq:totalEnergy:all:terms}, we arrive at the expression for the total energy of a BEC, Eq. \eqref{eq:totalEnergy_TF_approx}, valid within the TF regime.

\subsection{Angular momentum operator in original coordinates}
\label{sec:appAngularMomentumOperator:original:coordinates}

Here we establish a relation for the angular momentum operator with respect to the center of mass in original coordinates. 
We start with the displaced wave function $\psi_\Mlabel{D}(t,\rr)$, 
which is the solution of the GP equation \eqref{GPequationD} in the center-of-mass coordinates. 
In these coordinates the angular momentum operator $\hat\Mbf{L}_\Mlabel{D}$ is defined in the usual way by
\begin{equation}\label{eq:def:angular:momentum_D}
 \hat\Mbf{L}_\Mlabel{D}  = \hat\rr \times \bpp \,.
\end{equation}

With the help of the inverse of transformation \eqref{DefpsiD} we can rewrite the expectation value 
of the angular momentum operator \eqref{eq:def:angular:momentum_D} as follows
\begin{eqnarray}
\fl \ave{\hat\Mbf{L}_\Mlabel{D}}_{\psi_\Mlabel{D}} \equiv \frac{1}{N} \int\limits_{\Reals^3} \,\Mdiff^3 x \,\psi_\Mlabel{D}^*(t,\rr) \,\hat\Mbf{L}_\Mlabel{D} \,\psi_\Mlabel{D}(t,\rr)  \nonumber \\
\fl\qquad\qquad = \frac{1}{N} \int\limits_{\Reals^3} \,\Mdiff^3 x 
 \,\psi^*(t,\rr) \,\hat{D}\big(\calR(t),\calP(t)\big) \,\hat\Mbf{L}_\Mlabel{D} \, \hat{D}^\dagger\big(\calR(t),\calP(t)\big) \,\psi(t,\rr)  \nonumber \\
\fl\qquad\qquad = \frac{1}{N} \int\limits_{\Reals^3} \,\Mdiff^3 x
 \,\psi^*(t,\rr) \,\hat\Mbf{L} \,\psi(t,\rr) \equiv \ave{\hat\Mbf{L}}_{\psi(t)} \,, \label{eq:app:momentum:expect:def}
\end{eqnarray}
where $\hat{D}\big(\calR(t),\calP(t)\big)$ is the displacement operator which fulfills 
the relations \eqref{displacewavefunction} and \eqref{propertydisplace}. 
Hence, the angular momentum operator in original coordinates reads
\begin{equation}\label{eq:def:angular:momentum:operator}
 \hat\Mbf{L} = \big(\hat\rr-\calR(t)\big) \times \big(\bpp-\calP(t)\big)\,.
\end{equation}

\subsection{Angular momentum in the time-dependent Thomas-Fermi approximation}
\label{sec:appAngularMomentum:expect:value}
We now evaluate the expectation value, Eq. \eqref{eq:app:momentum:expect:def}, of the angular momentum operator defined by Eq. \eqref{eq:def:angular:momentum:operator} with respect to the approximate wave function, Eq. \eqref{eq:general:EffectiveTimeEvolTF}, valid in the time-dependent TF approximation. By making use of the coordinate transformation \eqref{eq:Ehrenfest_substitutions} we obtain the relation
\begin{eqnarray}
 \ave{\hat\Mbf{L}}_{\psi(t)} =& \ave{\left\lbrace \Lambda(t)\left[\hat\rr - \calR(0)\right] \right\rbrace 
  \times \left\lbrace m \,\diff{\Lambda}{t}\, \left[\hat\rr - \calR(0)\right] + \Lambda^{-\tp}(t)\, \,\bpp\right\rbrace}_{\psi_0} \nonumber \\
    & - \left\lbrace\Lambda(t) \left[\ave{\hat\rr}_{\psi_0} - \calR(0)\right]\right\rbrace \times \left\lbrace\Lambda^{-\tp}(t)\,\calP(0)\right\rbrace \,. 
    \label{eq:app:angular:expect:step:1}
\end{eqnarray}
Taking into account the initial condition, Eq. \eqref{eq:RP_initial_conditions}, for the center-of-mass position $\calR(0)$, the second term in Eq. \eqref{eq:app:angular:expect:step:1} vanishes and we arrive at Eq. \eqref{eq:angular:momentum:expectation:value}.

\subsection{Angular momentum for the Thomas-Fermi ground state}
\label{sec:appAngularMomentum:TF:ground-state}
When we consider the TF ground state $\phi_\Mlabel{TF}(\rr)$, Eq. \eqref{eq:initialTFwavefunc}, as the initial state $\psi(0,\rr)$ for the expectation value of the angular momentum matrix operator, Eq. \eqref{eq:angular:momentum:tensor:expect:value}, the last two terms vanish, that is
\begin{equation}\label{eq:app:angular:momentum:vanishing:terms}
 \fl \Lambda(t) \ave{\left[\hat\rr - \calR(0)\right] \otimes \bpp^\tp}_{\phi_\Mlabel{TF}} \Lambda^{-1}(t) \;-\; \Lambda^{-\tp}(t) \ave{\bpp \otimes \left[\hat\rr - \calR(0)\right]^\tp}_{\phi_\Mlabel{TF}} \Lambda^\tp(t) = 0 \,.
\end{equation}
This results from the fact that the contributions of these two terms cancel each other when the momentum operator is applied to $|\phi_\Mlabel{TF}(\rr)|$. In addition, the contributions due to the phase $\Mi\calP(0)\,\rr/\hbar$ of the TF ground state, Eq. \eqref{eq:initialTFwavefunc}, vanish in consequence of the initial condition, Eq. \eqref{eq:RP_initial_conditions}, for $\calR(0)$. What remains is the contribution of the first term in Eq. \eqref{eq:angular:momentum:tensor:expect:value}, which can be evaluated with the help of Eq. \eqref{eq:TFexpect:square:position:step2}. Hence, we obtain the expression
\begin{equation}\label{eq:app:angular:momentum:expect:step2}
 \ave{\hat{L}}_\Mlabel{TF} = \frac{2\,\mu_\Mlabel{TF}}{d+4} \left[\Lambda(t)\Omega^{-2}(0)\diff{\Lambda^\tp}{t} - \diff{\Lambda}{t}\Omega^{-2}(0) \Lambda^\tp(t) \right]
\end{equation}
for the expectation value of the angular momentum matrix operator within the TF regime. By applying the angular momentum matrix associated with the matrix differential equation as given by Eq. \eqref{eq:angular:momentum:matrix:Lambda} to \eqref{eq:app:angular:momentum:expect:step2}, we obtain Eq. \eqref{eq:angular:momentum:tensor:TF:ground:state}.

\section{Solutions of the matrix differential equation for isotropic traps}
\label{app:Lambda:DGL:isotropic}

In this appendix we find the solution of the non-linear differential equation \eqref{eq:lambda_dgl_scalar} 
in the implicit form given by Eq. \eqref{eq:integral-equation-lambda} valid for isotropic initial traps.
In the case of such a trap, that is $\omega_i(0) = \omega_0$, all scaling parameters $\lambda_i(t)$ 
are identical, $\lambda_i(t)=\lambda(t)$, and Eq. \eqref{eq:lambda_dgl_scalar} then reads
\begin{equation}
 \label{eq:app:lambda-dgl-isotropic}
 \diffz{\lambda}{t} = \frac{\omega_0^2}{\lambda^{d+1}}
\end{equation}
with the initial conditions 
\begin{equation}
 \label{eq:app:lambda-dgl-initial condition}
 \lambda(0) = 1 \qquad \left.\diff{\lambda}{t}\right|_{t=0} = 0,
\end{equation}
resulting from Eq. \eqref{eq:lambda_initial_scalar}.

By considering Eq. \eqref{eq:app:lambda-dgl-isotropic} as the classical Newton equation for the ``coordinate'' $\lambda$, 
we can integrate this equation with the integral of motion
\begin{equation}
 \label{eq:app:lambda-dgl-isotropic-energy}
 \frac{1}{2}\left(\frac{{\rm d}\lambda}{{\rm d}t}\right)^2+\frac{\omega_0^2}{d}\frac{1}{\lambda^d}=
 \frac{\omega_0^2}{d}\,,
\end{equation}
which plays the role of ``energy'' corresponding to Eq. \eqref{eq:app:lambda-dgl-isotropic}. 
Here we have used the initial conditions given by Eq. \eqref{eq:app:lambda-dgl-initial condition}.
After solving Eq. \eqref{eq:app:lambda-dgl-isotropic-energy} with respect to $d\lambda/dt$ and 
using again Eq. \eqref{eq:app:lambda-dgl-initial condition}, we obtain the implicit solution
\begin{equation}
 \int_{1}^{\lambda(t)}\Mdiff\tilde{\lambda}\,\frac{\tilde{\lambda}^{\frac{d}{2}}}{\sqrt{\tilde{\lambda}^{d}-1}} = 
 \sqrt{\frac{2}{d}}\;\omega_0\,t 
\end{equation}
of Eq. \eqref{eq:app:lambda-dgl-isotropic} for the function $\lambda(t)$.

\section*{References}

\bibliographystyle{dcu}

\bibliography{References_final_v3}

\end{document}